\documentstyle[12pt]{article}
\def\lsim{\mathrel{\rlap {\raise.5ex\hbox{$ < $}}
{\lower.5ex\hbox{$\sim$}}}}

\newcommand{\pr}{\paragraph{}}
\newcommand{\be}{\begin{equation}}
\newcommand{\ee}{\end{equation}}
\newcommand{\bea}{\begin{eqnarray}}
\newcommand{\nn}{\nonumber}
\newcommand{\eea}{\end{eqnarray}}
\newcommand{\nd}[1]{/\hspace{-0.6em} #1}
\newcommand{\nk}{\noindent}
\def\gappeq{\mathrel{\rlap {\raise.5ex\hbox{$>$}}
{\lower.5ex\hbox{$\sim$}}}}

\def\lappeq{\mathrel{\rlap{\raise.5ex\hbox{$<$}}
{\lower.5ex\hbox{$\sim$}}}}

\begin{document}

\begin{titlepage}
\begin{flushright}
CERN-TH.7269/94 \\
ENSLAPP-A-474/94 \\
CTP-TAMU-26/94 \\
ACT-08/94 \\
\end{flushright}
\begin{centering}
\vspace{.1in}
{\large {\bf Some Physical Aspects of Liouville String
Dynamics  }} \\
\vspace{.2in}
{\bf John Ellis$^{a,\diamond} $},
{\bf N.E. Mavromatos$^{b,\diamond}$} and
{\bf D.V. Nanopoulos}$^{a,c}$
\\
\vspace{.03in}
\vspace{.1in}
{\bf Abstract} \\
\vspace{.05in}
\end{centering}
{\small  We discuss some physical aspects of our Liouville
approach to non-critical strings, including the emergence
of a microscopic arrow of time, effective field theories
as classical ``pointer'' states in theory
space, $CPT$ violation and the
possible apparent non-conservation of angular momentum.
We also review the application of a phenomenological
parametrization of this formalism to the neutral kaon system.}
\vspace{0.2in}
\nk $^a$ Theory Division, CERN, CH-1211, Geneva, Switzerland,  \\
$^b$ Theoretical Physics Laboratory,
ENSLAPP,
Chemin de Bellevue,
Annecy-le-Vieux, France, and  \\
On leave from
S.E.R.C. Advanced Fellowship, University of Oxford, Dept. of Physics
(Theoretical Physics),
1 Keble Road, Oxford OX1 3NP, United Kingdom,   \\
$^{c}$ Center for
Theoretical Physics, Dept. of Physics,
Texas A \& M University, College Station, TX 77843-4242, USA
and Astroparticle Physics Group, Houston
Advanced Research Center (HARC), The Mitchell Campus,
Woodlands, TX 77381, USA. \\
$^{\diamond}$ {\bf Invited contributions } to {\it the
First International Conference
on `Phenomenology of Unification from Present to
Future', Rome, `La Sapienza' University, March 22-26 1994}

\vspace{0.01in}
\begin{flushleft}
CERN-TH.7269/94 \\
ENSLAPP-A-474/94 \\
CTP-TAMU-26/94 \\
ACT-08/94 \\
May 1994 \\
\end{flushleft}
\end{titlepage}
\newpage
\section{Introduction and Summary}
\pr
One of the most profound issues in microphysics
is a consistent formulation of gravity. The only
candidate we have for resolving this issue is
string theory, which is known \cite{mandelstam}
to be free of the ``trivial'' perturbative divergences that
beset quantum calculations in a fixed smooth space-time
background. Potentially far more profound problems
arise when one considers curved backgrounds with
horizons or singularities. Semiclassical calculations
in such backgrounds indicate that a pure quantum-mechanical
description cannot be maintained, as reflected in the
thermodynamic description  of macroscopic black holes,
with non-zero temperature and entropy \cite{bek}.
\pr
The question arises whether such a mixed quantum description
is also necessary at the microscopic level, once quantum
fluctuations in the space-time background and back-reaction of
particles on the metric are taken into account. Hawking \cite{hawk}
has argued that asymptotic scattering must be described in terms
of a linear operator $\nd{S}$ relating ``in'' and ``out'' density
matrices :
\be
  \rho _{out} = \nd{S} \rho _{in}
\label{proti}
\ee
that does not factorize as a product of $S$ and $S^\dagger$ matrix
elements. It has been argued \cite{ehns} that, if this is the case,
there should be a corresponding modification of the quantum Liouville
equation describing the time-evolution of the density matrix:
\be
 {\dot \rho } = i [ \rho, H ] + \nd{\delta H} \rho
\label{defterh}
\ee
This would cause pure states to evolve into mixed states, yielding
a loss of coherence associated with the loss of information
accross microscopic event horizons, that would be enhanced
for macroscopic systems \cite{emohn}.
\pr
Ordinary local quantum field theory is an incomplete guide to these
issues, but string theory appears to resolve them. The scattering of
particles off a
black hole in string theory is described by a well-defined
$S$ matrix, which reflects the existence of an infinite set of
local symmetries (and associated conserved charges) that interelate
(and characterize) different string states in the presence
of a black hole, as long as quantum fluctuations in the space-time
background and back-reaction of particles on the metric are ignored.
However, we have argued \cite{emnqm,emndollar}
that non-trivial modifications
(\ref{proti}) of the effective field theory description
of scattering and (\ref{defterh}) of the quantum Liouville
equation appear once these effects are taken into account.
We describe such effects using non-critical string theory
\cite{aben}, with time introduced as a renormalization
group flow variable \cite{emnqm}, associated with a covariant
world-sheet
scale
introduced via the Liouville field \cite{polch}.
We do not review here the basics of this approach,
which are
described elsewhere \cite{emndollar,emnerice,emnharc}.
However, we do discuss some physical aspects of this approach
that are particularly relevant to the focus of this
meeting.
\pr
One is a possible microscopic arrow of time \cite{emnqm,emndollar}.
As we discuss in sections 2 and 3, string theory satisfies the general
condition for the existence of an irreversible ``ageing''
variable, as a corollary of the existence of the
Zamolodchikov metric \cite{zam} in the space of couplings
of two-dimensional field theories on the world sheet.
The arrow of time emerges as a result of the stringy
symmetry-induced couplings of light particles to
massive, non-propagating solitonic states of the string
in a black-hole background \cite{emn1}. In this respect, we see
some similarity with ideas advocated by Penrose in the context
of local field theories in highly-curved space-times
\cite{pen}.
\pr
A second issue is the emergence of semiclassical ``pointer'' states
which describe specific low-energy effective quantum field theories
obtained from string theory. As we discuss in section 4, these
also appear as a result of the inevitable couplings to unobservable
non-propagating solitonic string states.
\pr
Then we discuss the violation of $CPT$ \cite{emncpt} and other
conservation laws in section 5. $CPT$ is expected to be violated
in any theory, such as ours, which allows pure
states to evolve into mixed states \cite{wald}.
As we have discussed elsewhere, energy \cite{emncpt} and
probability \cite{emnqm} are conserved in our string approach
to density matrix mechanics. However, the renormalizability
of the theory, which guarantees energy conservation, does not
guarantee the conservation of angular momentum. Unlike
string contributions to the increase in entropy, which cannot
cancel, the apparent non-conservation of angular momentum
may vanish in some backgrounds, though not in one
cosmological background that we study.
\pr
Finally, in section 6 we review the formalism \cite{ehns,emncpt,elmn}
for describing phenomenologically possible modifications of quantum
mechanics and violations of $CPT$ in the neutral kaon system,
which is also discussed here by Huet \cite{huetalk, huet}.
\pr
\section{A Primer in Non-Equilibrium Quantum
Statistical Mechanics}
\pr
One of the open problems of any local field theory
is how to incorporate an irreversible time variable.
In the way formulated so far, at a `microscopic'
or `fundamental constituents of matter' level, time
reversal symmetry is unbroken. The arrow of time
comes later, when one makes a `reduction'
of the degrees of freedom in order to describe the
observable world. In cosmology, for instance, the
arrow of time is believed to be induced by `integrating out'-
in a path-integral sense -  unobserved states hidden behind the
`particle horizons'. In a similar spirit, in
thermodynamics the arrow of time, or the `second law'
as it is commonly called, is induced by the `open-ness'
of the subsystem under consideration. Boltzmann's
second law of entropy increase implies the existence of a Lyapounov
function (entropy) for the subsystem, which increases with time,
whilst the entropy of the total system remains constant.
In the modern formulation of Boltzmann's law \cite{bek},
where entropy increase is associated with `information leakage '
from the open subsystem to the `environment', this entropy increase
implies an irreversible arrow of `time' \cite{wald}.
\pr
{}From the above discussion it becomes evident that the concept
of an  irreversible time  variable in statistical
mechanics
is different
from that of time in
Einstein's General Relativity Theory, where
it is a coordinate of the space-time manifold
that is completely reversible.
Misra and Prigogine \cite{mp} have discussed
this issue
formally
from a quantum statistical mechanics point of view by introducing
the concept of `measurable entities' into this framework.
They assumed the existence of a system
whose distribution function in phase-space $\rho (p,q,t)$
evolves reversibly under the flow of a time co-ordinate, $t$.
Such a flow is described by a unitary transformation
$U_t$:
\be
\rho (p,q,t) = U_t \rho (p,q, 0) U_t^{\dagger}
\label{1}
\ee
Reversible time translations are generated
by the Hamiltonian
${\cal H}$
of the system,
and the time evolution equation for $\rho$
has the familiar Liouville form
\be
\partial _t \rho =[\rho, L]_{PB}
\label{two}
\ee
where $L=e^{-i{\cal H} t}$ is the Liouvillean, and
$[ , ]_{PB}$  denotes Poisson brackets or commutators multiplied
by a factor $i/\hbar$, depending whether one considers
classical or quantum systems.
\pr
To introduce a time arrow in the above framework,
Misra and Prigogine \cite{mp}
assume that the physical states, i.e.
the ones that can be measured in an experiment,
have a phase-space
distribution function ${\tilde \rho}$ which
is related to $\rho $ via a {\it non-unitary}
transformation :
$\Lambda$
\be
{\tilde \rho} = \Lambda \rho
\label{three}
\ee
The transformation $\Lambda$
may be an invertible
map or a projection $P$. The evolution of ${\tilde \rho}$
is governed by a strongly irreversible Markov process
(s.i.m.p.), defined by the adjoint $W^*_t$
of a positive semigroup $W_t$ defined for $t \ge 0$:
\be
{\tilde \rho }(t) = W^*_t {\tilde \rho}(0)
\label{four}
\ee
with $W^*_t$ satisfying the interwining condition \cite{mp}
\be
\Lambda ^{-1} U_t  \Lambda = W^*_t  \qquad  or \qquad
P U_t P = W_t^*
\label{fourb}
\ee
and
\be
\partial _t {\tilde \rho} = \Phi (L)  {\tilde \rho}
\qquad : \qquad \Phi (L) = \Lambda L \Lambda ^{-1}~  or ~
PLP
\label{five}
\ee
It can be shown that (\ref{five}) implies the existence
of a Lyapounov function, if and only if the following condition
of star hermiticity is satisfied \cite{mp}
\be
i\Phi (L) = (i \Phi (L))^* \equiv \Phi ^\dagger (-L)
\label{six}
\ee
A particular case of an invertible $\Lambda$ transformation
is provided
under certain conditions
by a non-conservative force in the statistical
mechanics of open systems
\cite{ktorides}.
If $F_i$ is such a non-conservative force for a system
whose hamiltonian ${\cal H}$ is such that $\partial {\cal H}/
\partial p_i \ne 0$ as in a model of
dissipation, then
the evolution of the density matrix is described
by an equation of the form \cite{fronteau}
\be
\partial _t {\tilde \rho} = -\{{\tilde \rho}, {\cal H} \}_{PB}
+ \frac{\partial {\cal H}}{\partial p_i} {\cal G}_{ij}
\frac{\partial {\tilde \rho }}{\partial p_j}
\label{seven}
\ee
with ${\cal G}_{ij}[F_i]$ a matrix depending
on the non-conservative forces.
It has been shown in ref. \cite{ktorides}
that (\ref{seven}) satisfies
the star hermiticity condition (\ref{six}) if and only if
\be
{\cal G}_{ij} = {\cal G}_{ji} \qquad : \qquad
{\cal G}_{ij} \in {\bf R}
\label{eight}
\ee
In this way the theory of such
an open statistical system
is connected to the $\Lambda$-transformation theory of
reduction to a physical subspace.
\pr
The issue of demonstrating that time has
an arrow
is a bit more
subtle and requires more careful consideration. In the Misra-Prigogine
\cite{mp} approach, the starting point is a unitary transformation
$U_t$, which governs the time $t$ flow of the system $\rho$.
This implies straightforwardly that if one has a s.i.m.p.
in the positive time direction $t \ge 0$, described
by a
transformation
$\Lambda_{  + } $, then one will also have
a s.i.m.p. in the $t < 0$ direction governed by a
transformation
$\Lambda _{ - } $. The time arrow is
introduced into the system only if the dynamics
implies that \cite{mp} $ \Lambda _{ + } \ne \Lambda _{ - } $.
\pr
If the above conditions are met, the construction of an
internal time or age operator, $T$, is possible
according to a theorem of Misra and Prigogine \cite{mp}.
The operator $T$ is defined in terms of its eigenvalues $\lambda$
as follows:
\be
U_t T U_t ^\dagger = I + \lambda T
\label{nine}
\ee
and its action on physical states leads to a non-decreasing
sequence of the eigenvalues $\lambda$.
\pr
What we shall argue in this talk is that a similar
situation occurs in non-critical
string theory \cite{aben} when
microscopic quantum fluctuations of space-time
into
non-trivial,
highly-curved (singular) backgrounds are taken
into account \cite{emnqm} via non-critical
string theory. The existence
of such singularities in the structure of space-time
implies the existence of
non-decoupling,
delocalized
topological
string modes which are non-propagating and can be thought of
as remnants of a highly-symmetric (topological) phase of
string theory \cite{wittop, emntop}.
Such modes cannot be measured in local scattering experiments
conducted by a
conventional observer, and therefore define a sort
of `environment' for the propagating string modes.
Time is then introduced \cite{emnqm} into this framework
as a dynamical local renormalization group scale
on the world-sheet \cite{shore,osborn} (Liouville field),
and its arrow is established as a result of the unitarity
of the effective string
$\sigma$-model describing string propagation in the
singular background under consideration \cite{witt}.
We shall be very brief in this description due to lack of
space and time (!). The interested reader may find detailed
presentation of this work in ref. \cite{emnerice}
and references therein.
\section{The Two-dimensional Stringy Black Hole and the
Arrow of Time in String Theory}
\pr
In the first-quantized
$\sigma$-model string formalism
the propagation of string particles in non-trivial
backgrounds is described by deforming
the
$\sigma$-model Lagrangian by the appropriate vertex operators.
The couplings of such operators are the background fields
of the target space-time. Conformal invariance on the world sheet
implies the vanishing of the pertinent renormalization-group
$\beta$ functions, which are interpreted
as equations of motion for the backgrounds.
In this way the dynamics of any consistent
background is described by
marginal (in a renormalization-group sense) deformations
of the world-sheet action, that preserve by construction
the conformal symmetry. The construction of the
appropriate vertex operators for the various background fields
relies on the completeness of the set of operators with
zero anomalous dimensions, or $(1,1)$ operators.
As a result of this completeness property,
higher-level operators appear in the operator product expansion
(OPE) of lower-level operators. For instance,
starting from the lowest-level bosonic string states, the so-called
tachyons,
one finds the following OPE between the respective vertex operators
\cite{polybook}
\be
V_T(z,{\bar z}) \otimes V_T (z',{\bar z}') \propto
\frac{1}{|z-z'|^4} V_T (\frac{z + z'}{2},
\frac{{\bar z} + {\bar z}' }{2}) + \sum _{N \ge 1} V_N
(\rm less~ singular)
\label{ten}
\ee
where the sum is over higher-level string states $N\ge 1$,
and $V_N$ denote the associated vertex operators
(the corresponding less-singular world-sheet factors are
not exhibited for convenience).
\pr
The physical spectrum of the theory is defined in flat
space time, and is assumed background-independent,
for general covariance reasons. Thus the concept of higher-level
string states is defined unambiguously in any background.
These remarks will be crucial for our subsequent discussion
of the spectrum of the two-dimensional stringy black hole.
\pr
If the theory is formulated on a world sheet with fixed topology,
e.g. a sphere, then the corresponding background field theory
will be `classical' in target space. Quantum background
corrections are described by higher-genus effects on the world
sheet \cite{stringbook}. However, such effects can be
effectively projected onto the world-sheet sphere \cite{fischler},
by including corrections to the conformal invariance
conditions which supplement those
obtained by the perturbative renormalization-group
treatment on fixed-genus Riemann surfaces.
\pr
In this context, target-space quantum gravity effects,
such as black hole creation and evaporation, Hawking radiation, etc.,
can be studied by conformal field theory methods.
The only explicit example
which has been solved exactly from a conformal field theory
point of view is the two-dimensional
(target-space) black hole of ref. \cite{witt}. The model
is described by a gauged Wess-Zumino conformal field theory
on a non-compact group $SL(2,R)/U(1)$. This theory
posesses one propagating degree of freedom, a massless
scalar field called a `tachyon' for historic reasons,
and an infinity of non-propagating
higher-level string modes
with discrete energies and momenta \cite{pol}.
The target-space metric for this two-dimensional string theory
assumes the following
form in the Minkowski $SL(2,R)/O(1,1)$ case \cite{witt}:
\be
ds^2 = dr^2 - tanh^2 r dt ^2
\label{eleven}
\ee
where $r$ is space-like and $t$ is time-like. It has been shown
in ref. \cite{emnqm,emnharc} that, as a result
of the static nature of the background (\ref{eleven}),
one may consider it as a fixed point of a
renormalization group transformation
on the world sheet, with $t$ interpreted as
a local renormalization group scale. This
formalism is known as Liouville string
dynamics. The central charge of the model
is $c=\frac{3k}{k-2}-1$, with $k$ the Wess-Zumino
level parameter.
In this formalism,
due to the $c=26$ critical string character
in the case $k=9/4$,
one obtains dynamically a Minkowski signature
\cite{aben,polch,DDK,mm}
in target space-time, and
the Euclidean $SL(2,R)/U(1)$
counterpart
can be obtained by
analytic continuation \cite{emnharc}.
\pr
An important element in the stringy black hole (\ref{eleven})
is the existence of a dilaton field, whose presence is necessitated
by the non-critical dimensionality of the target space-time
\cite{aben}. From our point of view, the existence of a non-trivial
dilaton is important because it implies that a two-dimensional
black hole  can have a non-trivial entropy $S$
despite the absence of an
horizon area \cite{frolov}
\be
S = e^{\Phi _H}
\label{twelve}
\ee
where $\Phi _H $ is the value of the dilaton field at the
horizon point \cite{witt}. This helps justify
the use of a two-dimensional
toy  model as a prototype for realistic string computation
relevant to
the definition of time in string theory.
\pr
In the flat target space-time case of two-dimensional strings,
known as the $c=1$ string model, the tachyon deformation, which coincides
with the world-sheet cosmological constant operator,
is exactly marginal, i.e. its renormalization group $\beta$-function
vanishes identically. This implies the vanishing of the respective
OPE coefficients in (\ref{ten}). However, this is not so
in the
black hole case.
{}From the structure of the pertinent
OPE of two tachyon deformations \cite{chlyk},
one can immediately conclude that the exactly
marginal deformation that turns on a non-trivial
tachyon background turns on an infinity of higher-level
(topological) modes as well. In the two-dimensional
string theory one may use $SL(2,R)$-isospin quantum numbers
$(j,m,{\overline m})$, to classify
the states, where $m$ (${\overline m})$ are third components
of the isospin in left (right) sectors of the closed string,
characterized by a common $SL(2,R)$ isospin $j$. In
this framework, the exactly marginal deformation
of a Euclidean black hole\footnote{At present, exactly-marginal
deformations of Wess-Zumino black hole models have been
constructed only for the Euclidean case. There are technical
subtleties in the Minkowski case, which is assumed to be
obtained
by analytic continuation. See
ref. \cite{continu}
for a discussion of this
procedure.}
is represented \cite{chlyk}
by an infinite sum of vertex operators $W_{j,m,{\overline m}}$
\be
L_0^1 {\overline L}_o^1 = {\cal F}_{-\frac{1}{2},0,0}^{c,c}
+ W_{-1, 0,0} - {\overline W}_{-1,0,0} + \dots
\label{thirteen}
\ee
where
\be
{\cal F} ^{c,c}_{-\frac{1}{2},0,0}(r)
=\frac{1}{coshr}
F(\frac{1}{2},\frac{1}{2} ; 1, tanh^2r )
\label{fourteen}
\ee
with
\bea
&~&F(\frac{1}{2},\frac{1}{2},1;tanh^2r) \simeq
\frac{1}{\Gamma ^2(\frac{1}{2})}\sum_{n=0}^{\infty}
\frac{(\frac{1}{2})_n(\frac{1}{2})_n}{(n !)^2}[2\psi(n+1)-
2\psi(n+\frac{1}{2})+ \nn \\
&+&ln(1 + |w|^2)]
(\sqrt{1 + |w|^2}~)^{-n}
\label{wseven}
\eea
is the tachyon operator, transforming according to a
continuous $SL(2,R)$ representation,
denoted by the superscript $(c,c)$
\cite{chlyk}.
\pr
In a low-energy
world, all measurements are made by local scattering experiments
employing only the ${\cal F}_{-\frac{1}{2},0,0}^{c,c}$
deformation. The global modes $W_{-1,0,0}, \dots$ can
be measured by Aharonov-Bohm experiments \cite{emnhair}
in higher-dimensional theories (in which case the two-dimensional
string prototype considered above is viewed as describing
appropriate spherically-symmetric $s$-wave field configurations
\cite{emn4d}). Even in that case, the infinite number of
the associated quantum numbers makes a complete
measurement impossible in practice, and hence
the mere existence of these topological modes
implies a `quasiparticle' \cite{fronteau}
structure for the propagating string modes,
in the sense of modifying their energy-momentum dispersion
relations. This is a generic feature of a non-critical
string \cite{aben}. Any deviation from conformal invariance
in the sub-sytem of the propagating string modes will result
in Liouville-dependent dilaton terms. Such terms are
responsible for the existence of
screening charges in Liouville correlation functions.
Their presence affects the energy-momentum dispersion relation
of a propagating string mode. For instance, for the tachyon mode
one has, in the simplest case of a non-critical bosonic string,
of matter central charge $C_m$, propagating in a
linear dilaton background, \cite{aben}
\be
-\frac{1}{2}( E^2 + Q^2 -{\bf p}^2)=1   \qquad ; \qquad Q^2=C_m-25
\label{dispersion}
\ee
In our stringy space-time foam
case the modification is much more complicated, due to the
non-trivial structure of the background target space-time.
\pr
As we have discussed above, the coupling between
propagating
and
non-propagating string modes is a consequence of the world-sheet
conformal symmetry.
There is a deeper reason for this, which appears to be generic
in string theory. This is provided by
the infinite-dimensional gauge symmetries
that characterize strings and are responsible for level mixing.
Such symmetries can be
elevated from the world-sheet to target space-time
\cite{venez,evans}. Techniques for such an
elevation have been described
in the literature \cite{ovrut}, where we refer the interested reader.
For our purposes, we note that in the two-dimensional Wess-Zumino
strings
the world-sheet ancestors of the
target-space gauge symmetries are $W_\infty$ Lie-algebraic structures
pertaining to spin 2 and higher world-sheet operators \cite{emn1,bakir}.
The sub-alagebra generated by the
spin 2 (stress-tensor) operator is the conformal algebra, which
is part of $W_\infty$. It is a straightforward computation
to show that the world-sheet $W_\infty$ charges do not commute with the
string level number operator \cite{emnhair}. Hence the `innocent'
requirement of conformal invariance in the deformation
(\ref{thirteen}), which mixes the various string levels, finds
a natural explanation within the more general context of
stringy gauge
symmetries that have no analogue in any local field
theory. In higher-dimensional string theories,
such symmetries are spontaneously broken
by the graviton background field \cite{evovr}, as one expects
for a mass-level mixing symmetry in the presence of a non-zero
mass gap characterizing a propagating massive mode. However, in the
case of a topological mode the notion of a mass gap is not defined
in the same way.
Hence symmetries mixing propagating with topological modes
are not necessarily broken by non-trivial backgrounds. Moreover,
in two space-time dimensions spontaneous breaking cannot occur
\cite{coleman}, so in the case of the two-dimensional
black holes the
$W_\infty $ symmetries  should be considered as unbroken,
provided one takes the infinite set
of higher-level discrete modes
into account.
Their presence is essential for the maintenance of quantum
coherence \cite{emn1}, as a result of the phase-space area preserving
nature of such algebraic structures \cite{witt2}.
A low-energy measurement process effectively
breaks such symmetries,
since it effectively integrates out the discrete
modes in (\ref{thirteen}). This implies in turn the non-conservation
of the phase-space volume for the propagating modes under target-time
evolution, and hence a modification of the quantum mechanics for the
low-energy modes, which thereby behave as an `open' system
\cite{mp,fronteau}.
\pr
{}From this point of view, one can define a reduction to a
physically-measurable
subspace by a sort of $\Lambda$ transformation of the type
discussed by Misra and Prigogine \cite{mp} as follows
\cite{emnerice} :
\be
{\tilde \rho}(X,P,g^i,p_i, t) \equiv e^{F[g^i,p_i,t]-
{\cal H}[g^i,p_i,t]}
\label{fifteen}
\ee
where $t$ is introduced as a covariant
renormalization-group scale via a Liouville field,
$X$ and $P$ denote the phase space of the string particle,
and $g^i$,$p_i$ the background field theory phase-space over
which the string propagation takes place. In equation
(\ref{fifteen}),
${\cal H}$ is the corresponding 1st-quantized
string Hamiltonian \cite{stringbook}
and
$F$ is the $\sigma$-model
effective action that generates string amplitudes
for the propagating single-particle string modes $g^i$.
In the two-dimensional case these modes are just the tachyon
deformations ${\cal F}_{-\frac{1}{2}, 0, 0}^{c,c}$.
\pr
On the other hand, the `full' string theory, including the
global modes, which corresponds to the unitary initial
system $\rho$ of Misra and Prigogine \cite{mp}, is defined
through a $\sigma$-model deformed by the exactly-marginal
deformation (\ref{thirteen}):
\be
\rho(X,P,g^i,p_i, t) \equiv       e^{F[L_0^1 {\overline L}_0^1 , t] -
{\cal H}[L_0^1 {\overline L}_0^1, t]}
\label{sixteen}
\ee
The (abstract) analogue of a
$\Lambda$ transformation in this case is essentially
the process of a string path integration of the topological modes
in (\ref{sixteen}).
Formally such a procedure is still not rigorously known, given the
absence
of a satisfactory string field theory so far. In practice
however,
the first quantized approach, based on the $\sigma$-model description
(\ref{fifteen}), proves sufficient.
Using the previously-mentioned
theorem
of Misra and Prigogine \cite{mp},
one expects to be able to define an internal time,
that flows irreversibly under the dynamics of the subsystem
of the low-energy string modes. Indeed, in our case we can
see how to
define
an internal time by looking at the important difference
between (\ref{fifteen}) and (\ref{sixteen}). Equation
 (\ref{sixteen})
is mathematically consistent as it stands, given the
exactly-marginal nature of the deformation (\ref{thirteen}).
In this case, the local renormalization group scale
decouples from the background $\sigma$-model couplings,
which thus become static \cite{witt,emnharc}, and it only
appears as a coordinate of the target space of the $\sigma$-model.
In contrast, equation
(\ref{fifteen}) requires renormalization,
since the tachyon deformation (\ref{fourteen}) breaks conformal
symmetry, and hence induces ultraviolet divergences
in the world-sheet model. Consistency is restored by
appropriate Liouville dressing using the formalism
of curved-space renormalization in field theory \cite{shore,osborn}.
In our case,
to leading order in the inverse
Wess-Zumino level parameter $\frac{1}{k}$,
and hence to first order in the Regge slope $\alpha '$,
only the OPE coefficients in the $\beta$ function are important, and
hence
the appropriate Liouville dressing is given by \cite{emnharc}
\be
 \int d^2z g V_g(r) \rightarrow
\int d^2z ge^{\alpha _g \phi}V_g(r) \equiv
 \int d^2z g V_g(r) - \int d^2 g^2 C_{ggg}V_g(r)\phi + \dots
\label{seventeen}
\ee
where $r$ is a $\sigma$-model spatial coordinate, and
$g$ is a generic deformation coupling (for simplicity
we have assumed a single deformation).
It is conjectured
that a similar consistent result is obtained for finite $k$,
and hence to all orders in $\alpha '$.
\pr
The dressing (\ref{seventeen}) corresponds to
an effective anomalous dimension for the coupling
of the tachyon deformation, which depends on the zero mode
of the local renormalization scale/Liouville field $\phi _0$.
Technically, one employs the so-called
`fixed-area constraint' \cite{DDK} in correlation
functions by inserting the identity
\be
\int dA \delta (A- \int d^2z e^\phi \sqrt{{\hat \gamma}}) =1
\label{fixar}
\ee
where $A$ is a covariant world-sheet area, and ${\hat \gamma }$
denotes a fiducial world-sheet metric. On the one hand,
this formalism
allows
an integration over the Liouville mode $\phi$
in the $\sigma$-model
path-integral, as appropriate for its target-time
interpretation, whilst on the other it
preserves an
explicit global renormalization-scale $A$ dependence
in the integrand
of the $A$-integral corresponding to the `physically'
measurable part of the correlation functions, which thus
become target-time dependent.
The global scale corresponds to the zero mode of the Liouville
field, and this will always be understood in the following.
It will be essential for determining the physical sense
of the time flow to be discussed later.
It should be remarked
that in this formalism the renormalized
deformations of the $\sigma$-model Lagrangian
are conveniently written
as
\bea
\int d^2z g^i[X(z,{\bar z}),\phi (z,{\bar z})]
V_i[X(z,{\bar z}),\phi (z,{\bar z})] = \nn \\
\int d^2z\int d^Dy g^i(y) \delta^{(D-1)}({\underline
y} - {\underline X}(z,{\bar z}))\delta ^{(0)}(y^0- \phi(z,{\bar z}) )
V_i[X,\phi]
\label{renormdef}
\eea
where $D$ is the dimension of target space-time, including the
Liouville mode as a coordinate, and
appropriate normal ordering is understood.
In this way summation over the index
$i$ can include the integration over
space-time coordinates $y$. Equation (\ref{renormdef}) encapsulates
the relationship between the coordinate time $y^0$ and the
irreversible Liouville mode.
Thus, the above procedure
amounts to the advertized `temporal' dependence of the
$\sigma$-model background couplings, where the time appears
as
an `external'
irreversible evolution parameter in
target space-time and
is at the same time
related to
a non-trivial quantum world-sheet field (the Liouville mode) in the
$\sigma$-model.
In the two-dimensional black-hole example this
effect describes the collective effects of massive global string
modes (\ref{thirteen}).
\pr
The induced arrow of time
follows nicely from the properties of
correlation functions for tachyon (matter) deformations in
the
Liouville-dressed theory. The crucial assumption is that
the effective theory after integrating out
massive global string modes
is unitary on the world sheet. This implies that the induced
renormalization group flow as a result of the dressing (\ref{seventeen})
will be irreversible, as a consequence of the
Zamolodchikov C-theorem
\cite{zam}. The associated Lyapounov function is constructed
out of components of the world-sheet stress tensor
\be
C=2z^4 <T(\sigma) T(0)>-3 z ^3 {\bar z} <T_{z{\bar z}}
(\sigma) T(0) >
- 6 z^2 {\bar z}^2 <T_{z{\bar z}} (\sigma ) T_{z{\bar z}} (0) >
\label{eighteen}
\ee
where $T \equiv T_{zz}$, $T_{z{\bar z}} $ is
the trace of the stress tensor, and $< \dots >$
denotes a $\sigma$-model v.e.v. with respect
to the Liouville renormalized deformation $[{\cal F}_{-\frac{1}{2},0,0}
^{c,c}]$ in the sense
of equation
(\ref{seventeen}). The C-theorem states that
the function $C$ flows irreversibly under the flow of the scale
$t$, the zero mode of the Liouville field :
\be
\partial _t C =-12 <T_{z{\bar z}} (\sigma) T _{z{\bar z}}(0)> =
-12 \beta ^i <V_i V_j > \beta ^j  \le 0
\label{nineteen}
\ee
where the metric in coupling constant space
\be
 G_{ij} \equiv  2|z|^4
<V_i(z,{\bar z}) V_j(0,0) >
\label{zametric}
\ee
is positive
for unitary theories on the world-sheet.
\pr
A rather straightforward analysis yields
the following string analogue of the evolution equation
(\ref{seven}) \cite{emnqm}
\be
\partial _t {\tilde \rho} = -\{ {\tilde \rho}, H \}_{PB} +
\beta^j G_{ij} \frac{\partial}{\partial p_i }{\tilde \rho}
\label{strli}
\ee
where
in the two-dimensional
black-hole example the couplings $g^i$ denote the propagating
`tachyon' modes. The key point to notice here
is that this equation has the form (\ref{seven}),({\ref{eight})
which guarantees the existence of an ``ageing'' operator
in non-equilibrium quantum statistical mechanics, i.e.,
a microscopic arrow of time. This is a consequence
of the symmetry and reality properties of the Zamolodchikov
metric in coupling constant space for unitary world-sheet
theories \cite{zam}.
\pr
It may be useful to comment here on
the precise meaning
of the partial derivative with respect to the zero mode
of the renormalization scale.
As stated above, this denotes dependence on the covariant
world-sheet area $A$, which occurs in the fixed world-sheet area
expectation values $< \dots >$. The
integration over the Liouville field in the fixed-area-constraint
formalism (\ref{fixar}) implies that $C$ is a target space-time
action functional. For stringy $\sigma$-models it is known \cite{curci}
that any form of the $C$-theorem which is local in target space
will yield
anomalous-dimension-like terms
on the right-hand-side of (\ref{nineteen}),
which are due to world-sheet infrared divergences.
For
graviton
$G_{MN}$ and dilaton $\Phi$ backgrounds they assume the form
\be
\gamma (G,\Phi) = -\frac{3}{16\pi^2} \frac{\alpha '}{2}
 \nabla ^2 (\beta _\Phi - \beta_M^M +
\dots )
\label{anomalous}
\ee
where the $\dots$ indicate higher-order $\alpha '$-corrections.
Such anomalous dimension terms
spoil the monotonicity properties of the local $C$-function,
even for unitary theories.
Fortunately, due to the fact that they are total
target space-time derivatives,
they are absent\cite{mavc}
in the space-time integrated form of the $C$-theorem.
Thus, the positivity of the action functional (\ref{eighteen})
is guaranteed for unitary
theories, in which case the action $C$ counts
correctly the physical degrees
of freedom
of the system.
Moreover, this integrated form guarantees
the existence in the associated flow equations
of second order derivatives
with respect to the renormalization-group scale,
which arise from the usual functional derivatives
of a string effective action  with respect to the
background fileds/couplings \cite{emnqm,emnerice}.
However, the existence of (dissipative)
friction terms, as a result
of the non-critical nature of the underlying
conformal field-theory \cite{emnqm}, implies the
irreversibility of such a flow.
This property applies in a fixed-genus (sphere) computation.
In higher genera there will, in general, be mixing
of standard matter states with ghosts circulating along the
handles of the Riemann surface. However, measurement in
our
effective
theory cannot detect
such effects
directly.
Following the analysis of ref. \cite{fischler}, it is possible
to project higher-genus effects onto the lowest-genus Riemann
surface (sphere), in such a way that the string loop effects
appear as extra renormalization counterterms on the world-sheet
theory, not included in a perturbative fixed-genus
renormalization. Thus, in such a formalism the lowest-genus
effective theory can always be assumed unitary, and the effective
$C$-theorem is applicable \cite{espriu}.
\pr
When one resums over higher genera, the classical
couplings $g^i$ become `quantum operators' ${\hat g}^i$, and
the associated Poisson brackets become commutators.
The quantum version of (\ref{strli}) is \cite{emnqm}
\be
\partial _t {\tilde \rho} =i[{\tilde \rho}, {\cal H} ]
+ i \beta^i G_{ij} [g^j, {\tilde \rho}]
\label{twentythree}
\ee
\pr
{}From equation
(\ref{nineteen}) one can construct an associated
statistical
entropy which is essentially the exponential of $C$ (\ref{eighteen}).
It has been shown \cite{emnqm}
that this quantity can be expressed in terms of the
reduced density matrix ${\tilde \rho}$ of the low energy subsystem
according to Boltzmann's prescription
\be
S = e^{-C} = - Tr {\tilde \rho} ln{\tilde \rho}
\label{twenty}
\ee
which for unitary theories varies monotonically
along the renormalization group flow
\be
{\dot S} =   \beta^i G_{ij} \beta^j S \ge 0
\label{monotonicentr}
\ee
\pr
We have argued in ref. \cite{emndollar,emnharc}
that world-sheet instanton effects can be used
to describe qualitatively
both higher-genus and global mode effects
in the Wess-Zumino black hole model. A rigorous treatment
of this issue has not yet become available, and it probably has
to wait for a satisfactory matrix model
formalism \cite{matrix}
of black holes in string theory, where a summation
over higher genera is performed exactly\footnote{For some
attempts towards this direction see ref. \cite{bhmatrix},
where a matrix model scenario for string propagation
on a fixed black hole geometry is presented.}.
The  association of world-sheet instanton
with higher-genus effects is, however,  supported
by a study of the $N=2$ topological
Wess-Zumino theory on a black-hole space-time background.
Such models are believed \cite{emntop}
to describe correctly the topological
phase of the two-dimensional string, which
is also reflected in the
black hole singularity itself \cite{witt,eguchi}.
The connection with higher genera comes from the
conjecture of Mukhi and Vafa \cite{mukhi} that a $c=1$
string theory resummed over genera is expressed
as a topological world-sheet Wess-Zumino model
formulated on a $SL(2,R)/U(1)$ group manifold.
\pr
We are concerned next
with the sense of the flow of
the subsystem of the propagating string modes.
In ref. \cite{emndollar} we
have established a flow of time which is
opposite to the conventional renormalization group flow.
There is an unambiguous way of determining the correct sense of
the
flow,
associated with the fact that, according to Zamolodchikov,
there is an `thinning' of physical degrees of freedom
of the system  along the renormalization-group flow.
In unitary theories, where the original analysis
of Zamolodchikov took place \cite{zam}, the effective
central charge counts correctly such degrees of freedom.
This notion can probably be extended \cite{kutasov} even to
non-unitary theories. In our case, as we shall discuss below,
the `bounce' picture of Liouville flow \cite{kogan,emndollar}
clearly provides a mathematically rigorous construction which
selects
the direction of flow along which there is a `thinning '
of degrees of freedom in the observable world:
close to the starting point of the flow,
in the topological string phase, the delocalized modes
are strongly coupled to the propagating string modes.
On the other hand, at the end of the flow, where flat space-time
is approached asymptotically, the topological modes decouple.
The associated entropy production (\ref{twenty})
in such a picture, is interpreted as pertaining to
the amount of information carried by the `environment'
of the topological modes.
\pr
We now develop briefly the `bounce' picture.
Consider an $N$-point correlator of `tachyons' in the
above Liouville string. Its expression
is \cite{Li}
\be
<V_{i_1} \dots V_{i_N} >_\mu \propto (\int dA e^{-A} A^{-s-1})
<V_{i_1} \dots V_{i_N} >_{\mu =0}
\label{twentyone}
\ee
where the subscript $\mu$ denotes world-sheet cosmological constant
deformations, appropriately modified in the black hole where
$\mu$ is related to the black hole mass \cite{bershadsky}. The
quantity $s$ is a kinematic factor involving Liouville energies
of the various operators.
The quantity $A$ is the covariant area of the world-sheet that sets
a renormalization scale \cite{kutasov}, and has been introduced
via the fixed-area constraint (\ref{fixar}).
The $A$-integral is ultraviolet divergent if $s$ is a positive integer,
which is the case in a two-dimensional black hole where scattering
of tachyons off the black hole results in an excitation of the latter
to discrete string states \cite{emnhair}.
The regularization of the integral can be made by analytic
continuation,
representing it by the contour integral depicted in fig. 1
\cite{emndollar,emnerice}.
The contour of fig. 1 implies a `bounce' \cite{kogan,emndollar}
of the world-sheet (c.f. figure 2)
area at the infrared fixed point ($A \rightarrow \infty $)
towards the ultraviolet one ($A \rightarrow 0$).
\pr
The bounce interpretation of the Liouville renormalization
group is different from the ordinary representation
describing transitions among string vacua. The bounce picture is in
perfect agreement with the time reversal symmetry breaking
$\Lambda$-transfor-
mation approach of Misra and Prigogine \cite{mp},
if one identifies the two opposite directions of time
in fig. 1
with the $\Lambda _{\pm}$ branches.
The bounce picture is supported by explicit
instanton computations in the dilute-gas
approximation on the world sheet, as discussed in
\cite{emndollar,emnharc}.
\pr
\section{Quantum-Classical Correspondence within the
Renormalization Group Framework}
\pr
In the previous section we
have developed a deterministic
flow of time, which stems from working
within an effectively fixed-genus $\sigma$-model.
In this picture, higher genus effects are collectively
and qualitatively represented by world-sheet instanton
effects, that produce extra logarithmic scale
dependences not appearing within the conventional
fixed-genus renormalization group analysis.
However, from a formal point of view, summation over higher
genera leads to a natural quantization of the target-space
fields $g^i$, and one loses the concept of
a classical point in coupling constant phase space that evolves
determinstically.
To include this feature we seek
states in this quantum-mechanical
system that evolve `almost reversibly' in time, and therefore
are the closest quantum counterparts to the classical
points in the $\sigma$-model background phase space.
Given that the degree of irreversibility of an open
system is measured by the entropy production,
the above states should correspomd to minimum entropy
production \cite{partovi,zurek}. Their time evolution
follows classical phase-space trajectories, which
in our case express the usual renormalization-group flow.
In conventional quantum-mechanical treatments
these states are called `pointer states'.
For instance,
in the case of the harmonic oscillator, the
pointer states are identified with the
conventional `coherent states' \cite{zurek}.
The existence, as well as the nature,
of pointer states depends
on the form of the interaction of the open subsystem
with the environment \cite{albrecht}. For instance,
in a toy two-state spin system, it can be shown that
oscillatory pointer states appear at weak coupling with
the environment,
while
constant pointer states appear at very strong coupling.
For intermediate couplings there are no pointer states,
but only a noisy background \cite{albrecht}.
\pr
It is the purpose of this section to address these issues
for our system. As already mentioned,
a difference from the ordinary quantum-mechanical
case appears because
the `phase space' refers to
background fields $g^i$, and their conjugate momenta $p_i$
in the target space of the string, where the index $i$ includes
the target-space coordinates.
As a preliminary to finding our pointer states we follow
\cite{zurek} and
compute the linear entropy production:
\be
\partial _t s^{l} \equiv \partial _t (Tr{\tilde \rho} - Tr{\tilde \rho}
({\bf g},{\bf p},t)^2) = -\partial _t Tr {\tilde \rho}^2
\label{linentr}
\ee
in our framework.
Using equation
(\ref{twentythree}) it is straightforward to derive
\bea
&~&   \partial _t s^l =
i Tr ([\beta ^iG_{ij} , {\tilde \rho} ]g^j =\nn \\
 2&i&  Tr{\tilde \rho}^2 g^i \frac{d}{dt}G_{ij} g^j -
  2i Tr {\tilde \rho} g^i \frac{d}{dt} G_{ij} {\tilde \rho} g^j-
     2i Tr {\tilde \rho} g^i [ G_{ij} ,  {\tilde \rho} ]\beta^j
\label{linentrpr}
\eea
Using equation
(\ref{fifteen}), we can express the last term in terms
of the commutator
$[ G_{ij}, e^{-H} ] $, which is non-zero as a result
of the renormalization-group invariance
of $\beta^i G_{ij} \beta^j $, and the fact
that $\frac{d}{dt}\beta^i \ne 0$ in any scheme.
Choosing a scheme such that \cite{zam}  $G_{ij}=\delta_{ij} + O(g^2) $,
we observe that,
for the case of $(1,1)$ deformations that  we are interested in,
the last term is necessarily of order $O[g^6]$ and it is not
dominant in the weak-field $g^i << 1$ approximation, which we assume
for convenience.
The remaining terms can then be time-integrated to yield
\be
s^l(t) = i \int^t _{t_0} dt (Tr {\tilde \rho}^2 g^i  \frac{d}{dt}
G_{ij} g^j -i Tr{\tilde \rho}g^i \frac{d}{dt} G_{ij} {\tilde \rho}
g^j)
\label{slin}
\ee
Writing  $i\frac{d}{dt} G_{ij}$ as $-[G_{ij} , H ]$,
and taking into account the fact
that in the weak-field limit
the renormalization of $G_{ij}$ requires linear subtractions
in the logarithmic scale $t=ln\mu$, one may make the replacement
\be
[G_{ij}, H] \propto -\delta _{ij}  + \dots
\label{replace}
\ee
where the $\dots$ indicate higher order corrections.
Thus, the final result for the linear entropy produced in the time
interval
$2\Delta t  $ is
\be
  s^l (2\Delta t) = \int _{t-\Delta t}^{t + \Delta t} dt (
Tr {\tilde \rho}^2 g^i \delta _{ij} g^j -
Tr{\tilde \rho}g^i
\delta _{ij} Tr{\tilde \rho}g^j )
(\Delta g)^2
\label{finres}
\ee
Were it not for the self-interaction Hamiltonian
of the low-energy string-mode system, the natural candidates
for the pointer states would be the position eigenstates
in coupling-constant space, corresponding to background fields
in the target space of the string.
The non-triviality of the Hamiltonian
modifies this result by introducing `momentum' uncertainties
in coupling constant space. To see this, we may rewrite
the right-hand-side of (\ref{finres}) in the twin limit of
weak field and weak
coupling with the environment, as
\be
s^l \simeq \int _{2\Delta t}
dt ( <g^i \delta _{ij} g^j > - <g^i>\delta _{ij} <g^j>)
=\int _{2 \Delta t} dt <( g^i    - <g^i>)^2>
\label{100}
\ee
For the interaction Hamiltonian we use the
matrix-model inverted harmonic oscillator approximate
Hamiltonian \cite{matrix}
which is supposed to describe in a closed form the
result of the resummation over world-sheet genera for a two-dimensional
string,
\be
H=\frac{1}{2}(p^2 - q^2)
\label{101}
\ee
where $q$, $p$ are related by a
canonical
collective coordinate
transformation to the matter tachyon field of the two-dimensional
string theory and its
conjugate momentum respectively \cite{matrix}.
{}From our point of view, this change of variables corresponds
to a renormalization group choice, and hence our previous
analysis applies. In particular, we can use the known
solution of the inverted harmonic oscillator problem
to write (\ref{100}) in the form
\be
s^l (2\Delta t)= \int _{2\Delta t}  <\psi |
[(q - <q>)cosht + (p - <p>)sinh t]^2 |\psi >
\label{harmonic}
\ee
where $|\psi >$ denote the states of the inverted
harmonic oscillator. It is understood that any modification
of the matrix model potential, e.g., of the form
-$\frac{m^2}{q^2}$
which describes \cite{bhmatrix}
tachyon propagation in the background of a string
black hole of mass $m$, has been omitted here, since
such terms have been integrated out by the measurement
process which `sees' only the propagating matter.
The effect of the black hole is the entropy production
due to the non-zero coupling with the topological modes of the
string (of which one is the two-dimensional
space-time graviton itself).
In this specific model one can explicitly verify (\ref{replace})
as follows: as already commented, we can interpret
the `coordinate' $q$ and the conjugate
`momentum' $p$ as a specific scheme choice. Unitarity of the $\sigma$
model theory requires the following generic form
for the metric $G_{qq}$:
\be
G_{qq} = (const)^2  + (\alpha)^2 q^2 + \dots \quad ; \qquad \alpha
\in {\bf R}
\label{metricq}
\ee
Using equation (\ref{101}),
the canonical commutation relation $[q,p]=i\hbar$, and
the representation  $p=-i\hbar \frac{\partial}{\partial q} $,
it is straightforward to arrive at
\be
[G_{qq}, H] =-\alpha^2 + O[q\frac{\partial}{\partial q}]
\label{teliko}
\ee
thus verifying equation
(\ref{replace}).
\pr
We now remark that, for
the inverted harmonic oscillator case, $2\Delta t$
can be taken to be the infinite interval $[-\Lambda, \Lambda ]$,
with $\Lambda \rightarrow \infty $.
The result of the time integration then is proportional to
\be
   s^l \propto [(\Delta g)^2  + (\Delta p)^2 ]
\label{formula}
\ee
which is our final result that can be compared with the
conventional harmonic oscillator case \cite{zurek}.
The pointer states are found by minimizing  the entropy $s^l$.
Viewing the latter as a functional of $(\Delta g)^2$ and of the
product $\Delta g^i \Delta p_i \ge \hbar /2 $ (no sum over $i$ ),
and minimizing (\ref{formula}) with respect to the $\Delta g^i$,
we find that the states of minimum entropy are characterized
by
\be
  \Delta g^i \Delta g_i = \frac{\hbar}{2}
\label{pointer}
\ee
which is similar to the result of ref. \cite{zurek}, showing
that the minimum-entropy-producing initial states
are minimum-uncertainty Gaussian wave-packets of the
 inverted
harmonic oscillator that describes the dynamics of the
(quantum) two-dimensional string.
\pr
In the case of the conventional harmonic oscillator,
such Gaussian wave-packets coincide with the Wigner
coherent states. However, in the case of the inverted harmonic
oscilaltor, a thorough analysis \cite{barton}
shows that Gaussian wave-packets are different from
the Wigner coherent states. The latter are not
square-integrable, though still localizable because they fall like
$q^{-1}$ for $q \rightarrow \infty$,
and can be obtained by the action
of the Weyl operators on energy eigenstates \cite{barton}.
On the contrary, the Gaussian wave-packet assumes the usual
form
\be
    \psi (q, t=0) = (b\sqrt{\pi })^{-\frac{1}{2}} exp(-q^2/2b^2)
\label{gaussian}
\ee
An interesting property of equation (\ref{gaussian})
is that the corresponding time-dependent probability
distribution $|\psi (q,t)|^2$
retains its Gaussian shape under time evolution.
This result can be
derived from
the action of the
Green function of the inverted harmonic oscillator
\cite{barton} on equation (\ref{gaussian}). The
time evolution of the probability distribution
yields a time-dependent width
\be
{\tilde b}^2(t) = b^2cosh^2 t + b^{-2} sinh^2 t
\label{time}
\ee
We can also consider the scattering of a Gaussian
wave-packet incident on the inverted oscillator potential,
i.e., a Gaussian distribution in both $q$ and $p$.
As the initial state at $t=0$ we take \cite{barton}
\be
\psi (q, 0)=(b\sqrt{\pi})^{- \frac{1}{2}}
exp(-(q-q_0)^2/2b^2 + i p_0 x)
\label{xpg}
\ee
Its time evolution is found again with the help of the
appropriate Green function, and the result for the
probability density
is
\be
|\psi (q, t)|^2 = \frac{1}{\sqrt{\pi}{\tilde b}(t)}
exp(-(q -q(t))^2/{\tilde b}(t)^2)
\label{prob}
\ee
where ${\tilde b(t)}$ is given by (\ref{time}),
and $q(t) = q_0cosht + p_0sinht $ is the classical
trajectory of a particle having energy
${\cal E}=\frac{1}{2}(p_0^2-q_0^2)$. Notice that
the mean energy $<\psi | H | \psi > ={\cal E}-(b^2-1/b^2)/4 $.
It coincides with the energy ${\cal E}$ for minimum-uncertainty
wavepackets with $b=1$ (in units of $\hbar = 1$), which is the case
of the pointer states under consideration.
It is straightforward to see that the peak of the
Gaussian distribution (\ref{prob}) follows
the classical trajectory exactly \cite{barton}.
\pr
A more interesting feature that is directly relevant
to our Markovian approach to Liouville string
is the effect of the environment, simulated by
oscillators \`a la Caldeira and Leggett \cite{cald},
on the shape of the Gaussian distribution
(\ref{gaussian}), viewed as an initial pure state
of the sub-system at $t=0$. Assuming thermal
equilibrium with the heat bath of the
environmental oscillators, the authors of
ref. \cite{cald} simulated
the coupling to the environment
by a temperature-dependent fluctuating non-conservatve force
$F(t,T)$
\be
<<F(t,T) F(t', T)>> = 2\eta  k_B T  \delta (t -t ')
\label{fluct}
\ee
where $<< \dots >> $ denotes a statistical/thermal average,
$ t$ denotes the time,
$k_B$ is Boltzmann's constant,
and $\eta$ is a friction coefficient.
The result of ref. \cite{barton}
is that under the influence of the heat-bath
a pure initial state (\ref{gaussian})
evolves at time $t$ to a probability distribution
for the particle, described by the diagonal element
of the density matrix, which is given
exactly by the
Gaussian \cite{barton}
\be
\rho (q,q,t)= \frac{1}{\sqrt{\pi}   {\tilde b}(t,T)}
exp(-\frac{q^2}{{\tilde b(t,T)}^2})
\label{tempgauss}
\ee
where for a unit-mass and -frequency oscillator
\bea
&~&
{\tilde b}^2(t,T)=(
\frac{2 sinh\Omega t}{\Omega e^{\frac{\eta}{2} t}})^2
\{ (\frac{b}{2})^2 ( \Omega coth\Omega t + \frac{\eta}{2} )
+ (\frac{\hbar }{2b})^2  + \nn \\
&~& \frac{\hbar \eta}{2\pi}
\int _0^{\nu_{max}} d\nu \nu coth (\frac{\hbar \nu}
{2k_BT}) |\frac{e^{\frac{\eta}{2} t}}{sinh \Omega t} \int _0^t
d\tau e^{-\frac{\eta}{2} \tau + i \nu \tau } sinh\Omega \tau |^2 \}
= 2<<q^2 >>
\label{width}
\eea
with $\Omega \equiv \sqrt{1 + \eta^2/4}$.
\pr
What is the meaning of
 this result in our framework?
To answer this question it is
necessary to define a notion of ``fine'' thermodynamics
in the sense of Fronteau \cite{fronteau}, which allows the
definition of a phase-space-dependent `temperature'
$T^f(q,p,t)$ even for systems outside thermal equilibrium.
In the example considered by Fronteau \cite{fronteau}
the work done by a non-conservative force $F$ acting on the
system is given infinitesimally by
\be
dQ^f = F dq
\label{work}
\ee
and is related to the ``fine'' entropy $S^f (p,q,t)$ and
the ``fine'' temperature $T^f
(q,p,t)$ by the conventional
thermodynamic relation, in infinitesimal form
\be
dS^f (q,p,t) =\frac{dQ^f (q,p,t)}{T^f (q,p,t)}
\label{fine}
\ee
{}From section 2 and ref. \cite{emnerice}  we recall that
in our case the non-conservative forces are given by
\be
 F_i (q) = G_{ij}\beta^j
\label{noncons}
\ee
where $G_{ij}$ is the Zamolodchikov metric (\ref{zametric})
in $\sigma$-model
coupling constant space.   Using equation
(\ref{monotonicentr}),
and writing (\ref{fine}) in terms of differential rates,
we can define a ``fine'' temperature in our case
\be
   T^f ({\bf g},{\bf p},t) \equiv S^{-1} = e^{   C_{eff}({\bf g},t)
    - 25 }
\label{temperature}
\ee
with $C({\bf g}, t)$ the Zamolodchikov  $C$-function (\ref{eighteen}),
and the critical value $25$ arises from an appropriate choice of
normalization conventions. In this way, a notion of ``fine''
temperature is defined in non-critical strings, which
measures the deviation from conformal fixed points.
Near the infrared fixed point on the world-sheet, where $C_{eff}
\rightarrow \infty$ \cite{emnerice,emnharc}, the topological
phase of the string
is approached \cite{emntop},
and the ``fine'' temperature is infinite.
Adapting the formalism of ref. \cite{hu},
we find that the uncertainty
in
$\sigma$-model
coupling constant/field space diverges in this limit. This implies
the breakdown of
a low-energy point-like field theory, which should
be expected in the topological phase.
On the other hand, in the ultraviolet limit  $C_{eff} \rightarrow
25$ and the theory approaches that of a conventional
critical string.
In this case the unit of ``fine'' temperature is reached, and
`thermal equilibrium' is achieved. This is in agreement
with the law (\ref{fine}), because in this case,
both $\delta Q ^f$ and $\delta S^f$ vanish, since
they are
related to
temporal changes in momentum and position in phase-space,
whilst
$T^f$ is
still finite for a free
`particle' with momentum $p$,
being related to its kinetic energy \cite{fronteau}
$\frac{3}{2} T^f =p^2/2m$.
We note at this stage that the above definition of
temperature as a deviation from conformal equilibrium
in non-critical strings is not new. It has appeared
previously in the literature \cite{thomas}, when
topological defects on the world-sheet were considered.
Our ``fine'' temperature, which is a quantity
defined in coupling-constant space,
is related to the world-sheet
temperature of ref. \cite{thomas} by a simple exponentiation,
  \be
    T^f =
     e^{     C_{eff} - 25} \equiv
    e^{ T_{ws}}
\label{wst}
\ee
The physical reason behind this relation
is the requirement
of vanishing fine entropy and heat-transfer in the critical string, as
explained above.
\pr
Having defined a fine temperature, we can now interpret
the result (\ref{tempgauss},\ref{width}) in our framework,
by first replacing the equilibrium temperature $T$
by the fine quantity $T^f$ (\ref{temperature}).
Secondly, we take
into account the results of ref. \cite{bhmatrix}, according
to which the spatial coordinate $q$ of the inverted harmonic oscillator
potential can be related by an appropriate
canonical
collective coordinate
transformation to the
massless propagating
`tachyon'
mode of the two-dimensional string theory. From our point of
view, such a transformation corresponds to a renormalization-scheme
change, i.e., an appropriate coordinate change in the coupling
constant space of the two-dimensional string.
Thirdly, we recall that tachyon propagation
in a two-dimensional black-hole geometry is believed
to be described by a
modification of
the inverted-oscillator potential
by anharmonic terms \cite{bhmatrix}
\be
\delta V = - \frac{m^2} {2q^2}  + \dots
\label{pot}
\ee
describing the topological string modes,
where the  $\dots$ denote (an infinity of)
possible terms.
Integrating out such terms provides a low-energy
`observer'
measuring only the localizable
modes  with the analogue
of an `environment', where the
above decoherence results emerge.
Given that the minimum-entropy-producing initial states
are Gaussian, whose nature is not affected by the environment,
one can safely interpret the result  (\ref{tempgauss},{\ref{width})
as implying that the concept of a low-energy (observable) Gaussian
field theory mode survives the procedure of
integrating out the topological
string modes in the two-dimensional black-hole string theory.
The Gaussian wave-packet (\ref{tempgauss}) maintains
its minimum uncertainty in  coupling-constant space.
This should not be confused with the time-dependent
modified uncertainty in position and momentum in target-space
of a test string found in \cite{emnharc} as a result
of a time-varying minimum string length \cite{ciaf}.
In the field space of the stringy $\sigma$-model, the
uncertainty in measurements of a field and its conjugate
momentum, both related to pointer states in coupling-constant-space
quantum mechanics,
retains its conventional
form as in
local field theory,
unrenormalized by the Liouville mode.
\pr
It should be stressed that the above result, i.e., the
emergence of a quantum field theory as the low-energy
limit of a matrix model (or, in more general terms, of
a resummed world-sheet
$\sigma$-model theory), finds
a consistent explanation in
the formalism of
coupling-constant density-matrix mechanics.
In the simple Drude-model analogue example of ref. \cite{emnqm},
which is argued to capture the essential physical features
of the realistic string situation, it was been found that
the off-diagonal terms of the density matrix in coupling
space of the $\sigma$-model, $\rho (g^i, g^j, t)$,
behave like
\be
    \rho (g^i, g^j) \propto e^{-Dt (g^i-g^j)^2 + \dots }
\label{offdiag}
\ee
where $D$ is a small coefficient, depending on the
squares of the anomalous dimensions $\alpha _g$ (\ref{seventeen})
of the Liouville-dressed deformations $g_R^i$.
In our framework, the topological string modes, which have been
integrated out by the low-energy observer,
play the r\^ole of a continuously measuring
`apparatus', and the observed time flow is a result of such a
`measurement' process.
The `collapse of the wave-function', i.e.
the vanishing of the off-diagonal terms (\ref{offdiag})
of the density
matrix in coupling-constant space, occurs at $t \rightarrow \infty$,
which in our framework is the ultraviolet fixed point.
At that point critical string theory is recovered and a
fixed string background is achieved, in the sense
of a non-vanishing entry of only one of the diagonal
elements of the density matrix\footnote{It should be understood
that this picture would be modified by the action
of exactly-marginal deformations that induce uncertainties
in a specific background choice. This  is related to the well-known
string-vacuum degeneracy problem.
It is hoped that truly non-perturbative string effects
will eventually lift this degeneracy, leading to a unique
critical string vacuum, and hence to a single non-zero entry in the
string density matrix.}.
Away from this
equilibrium point
there are non-trivial interference terms (\ref{offdiag}),
expressing the quantum nature in the coupling-constant space
of the non-equilibrium physics. In this picture, critical string
theory is identified with the final result ($t \rightarrow \infty$)
of the
`measurement process' induced by the topological modes of the string
\footnote{This measurement process and the associated `collapse of
the
wavefunction' phenomenon in string coupling-constant space
should not be confused
with their analogues in the target space of the string, discussed
in ref. \cite{emnqm}. In that case one considers a density matrix
for a test string propagating in the non-trivial background $g^i$.
Its off-diagonal elements refer to spatially-separated points
in target space of the string, related to the above formalism
in the way explained in ref. \cite{emndollar}. In that case,
when many string
particles are present, the exponent of the
off-diagonal elements (\ref{offdiag}) acquires a multiplicative factor
$N$, the number of test strings, and the collapse time is diminished
significantly \cite{emohn}.}.
\pr
The minimum-entropy-production
 nature of the Gaussian wave-packet of the inverted
harmonic oscillator implies an almost reversible, deterministic
trajectory in coupling constant space for the resulting
states/fields away from equilibrium.
This combines
the classical deterministic nature of the renormalization group flow
with the quantum nature of the string backgrounds in the matrix model
formalism, that resums Riemann surfaces.
In the above example, as well as
in that of ref. \cite{zurek}, such states arise naturally through
decoherence effects associated with the interaction of the quantum
system with an `environment'. In the case of the black hole model,
as we discussed above,
the environment is provided by the topological modes of the string,
which do not decouple, and whose interaction with the propagating
low-energy modes results in an irreversible flow of time.
The pointer states  that arise through decoherence effects
in the coupling constant space of the stringy $\sigma$-model
move almost reversibly with the renormalization group time,
and in this way
one obtains a
conventional quantum field theory ($g^i$'s and $p_i$'s ) in the
target space of the string.
\pr
This demonstration in the inverted harmonic oscillator
approach appears to possess a deeper interpretation.
We remind the reader that it was this model that the geometric
interpretation of the $W_\infty$ string symmetry as a
coherence-preserving target-space symmetry was given \cite{emn1,witt2}.
The linear entropy is related to the area of the $p$ and $q$ phase space
by \cite{zurek}
\be
s^l =1- 1/A
\label{area}
\ee
for Gaussian distributions, like the ones corresponding to
our pointer states.
In the absence of black holes in space-time, the $W_{\infty}$
symmetries of the matrix model guarantee the invariance of the
two-dimensional phase-space area $A$ under (target)
time evolution\cite{emn1,witt2}.
Its time dependence (\ref{linentrpr}) indicates the breaking
of coherence as expected from the fact that the $W_\infty$
symmetries mix the propagating low-energy
string modes with the higher-level string states
which are delocalized
and do not decouple in the presence of a black hole \cite{emn1,emnqm}.
For the pointer states found above
this breaking of coherence is the softest possible one.
This justifies the use of the flat space-time
matrix model in writing down the temporal evolution
of the fields
$g^i$
in the weak coupling approximation.
\pr
The pointer states
are the closest approximations to classical points in
the string coupling-constant phase space. Due to their minimal
irreversibility, they are the best approximations
to the effective quantum field theories
usually taken as the low-energy
limits of string theories.
We note that the statistical entropy
(\ref{twenty}) is not minimized by the pointer states, as
can be easily seen by the fact that the linear entropy
$s^l $ (\ref{linentr}) provides only a {\it lower bound}
on the statistical entropy (\ref{twenty})\cite{elze}.
This implies an overall cosmological time arrow, whilst
in parallel allowing for the emergence
of almost-time-reversible local field theory
structures, associated with decoherence-induced
pointer states in the coupling-constant space.
In this way, we have a rather elegant way of understanding
how quantum field theory in target space arises in Liouville strings.
An interesting question is whether this is merely an elegant
formalism, or has some observable microscopic
consequences.
This may be possible,
as we argue
in the following section.

\section{Generic CPT Violation and Non-Conservation
of Angular Momentum}
\pr
The above definition of time in string quantum gravity made
it necessary to introduce non-perturbative effects
on the world sheet,
such as instantons. In their presence, certain
charges cease to be conserved, as a result of logarithmic
renormalization scale dependences. Such a situation implies
the non-commutativity of the Hamiltonian operator with the
respective charge operators on the world sheet. If one defines
a generalized $CPT$ symmetry  in such a way that
this symmetry leaves the mass of a string state invariant, but
changes the sign of the charge, then it is straightforward
to argue that in our case the elevation
of $CPT$ symmetry to target space fails in general. This
is a heuristic
argument, and one should really construct a rigorous proof
of such a stituation, which is not yet available.
\pr
However, the evolution of pure
states into mixed ones, as a result of the entropy
increase (\ref{twenty}) along the positive direction of time,  implies
the violation of $CPT$ symmetry as we now review in the
context of a general analysis
\cite{wald}.
Let us make the assumption that a $CPT$ operator ${\hat \Theta }$
exists
in target space, such that
\bea
\rho _{in}'&=& {\hat \Theta }^{-1} \rho _{out}  \nn \\
\rho _{out}'&=&{\hat  \Theta } \rho _{in}
\label{twentyonea}
\eea
and the $in$ and $out$ density matrices are related
through the superscattering operator $\nd{S}$
\cite{hawk}
\bea
\rho _{out} &=& \nd{S} \rho _{in}  \nn \\
\rho _{out}' &=& \nd{S} \rho _{in}'  \nn \\
\label{twentyoneb}
\eea
The following relation is a trivial consequence
of the above equations
\be
{\hat \Theta}  = \nd{S} {\hat \Theta} ^{-1} \nd{S}
\label{twentyfive}
\ee
which implies that $\nd{S}$ has an inverse. Clearly this
cannot happen if there
is evolution of
pure states into mixed ones and not vice versa, as implied
by the monotonic increase of the entropy (\ref{twenty}).
This proves the breaking of $CPT$ symmetry in the above framework.
\pr
In ref. \cite{emndollar} we have described non-factorisable
(i.e. $\nd{S} \ne S S^\dagger $, where $S$ is
the conventional $S$-matrix
operator) contributions to the string $\nd{S}$-matrix, coming
from valleys between topological defects on the world sheet
\footnote{Notice that, in that picture, the creation and
annihilation of
a target-space
black hole is represented as a world-sheet
monopole-anti-monopole pair \cite{emndua}. Instantons induce
transitions among such configurations of different charge,
the latter being proportional to the black hole mass. We note that
there is a formal analogy \cite{emndollar,emnerice}
with the Quantum Hall fluids: in that case,
instantons in the respective Wess-Zumino models, describing
the effective theories in `conductivity space',
induce transitions among the transverse-conductivity
plateaux \cite{pruisk}.}.
This provides an explicit demonstration
of the non-existence of an inverse $\nd{S}^{-1}$, and
hence of induced $CPT$ violation in the target space of
this effective string theory.
It should be stressed, however, that the above considerations
cannot exclude the possibility of a some weaker form of
$CPT$ invariance
\cite{wald} which might cause violations
of $CPT$ symmetry to be unobservable
in an experimental apparatus.
Such a situation falls beyond the scope of the present
talk, and in what follows we simply explore the
possibility that a
detectable violation of $CPT$ occurs, which we
parametrize in a way suitable for
present experiments with neutral kaons
and at future $\phi$ factories \cite{cplear,dafne}, that
constitute the most
sensitive
probes in a search for violations of quantum
mechanics at the microscopic level.
For more details we refer the reader to the
literature \cite{emncpt,elmn,huet} and also to
Huet's talk at this meeting \cite{huetalk}.
\pr
Before proceeding with the
parametrization of such possible phenomenological
effects,
it is worth pointing out two important properties
of our modification of
quantum mechanics due to stringy
quantum gravity effects.
The first is energy conservation on the average,
which follows from renormalizability of the world-sheet theory,
and the second is a generic violation
of angular momentum conservation.
Both properties can easily be understood formally
as follows.
The renormalized
background couplings $g^i$ are assumed to be quantum
operators, as a result of a higher-genus
resummation. This implies that the modified
density matrix equation (\ref{twentythree})
that describes the time evolution
will be used \cite{emnqm}.
\pr
Consider an operator {\cal K} in this framework
whose average is given by
\be
<<{\cal K}>> \equiv Tr ({\tilde \rho} {\cal K} )
\label{twentyfour}
\ee
Its time evolution is given by
\be
\partial _t <<{\cal K}>> = <<\beta^i G_{ij} [g^j, {\cal K}]>> +
<< \partial _t {\cal K} >>
\label{twentyfiveb}
\ee
where we have used the fact that $\beta^i G_{ij} $ is a functional
of $g^i$ only, and not of $p_i=\frac{\delta}{\delta g^i}$.
If the operator ${\cal K}$ is the $\sigma$-model Hamiltonian
(energy) then, using $[g^i, {\cal H} ]=\beta^i $
as well as the fact that the $C$-function is related
to the string effective action that
generates string amplitudes,
it is straightforward to derive \cite{emncpt,emnharc}
\be
\partial _t <<{\cal H}>> =\partial _t (\frac{\delta \beta^i}{\delta
g^i} ) =0
\label{twentysix}
\ee
as a result of the renormalizability of the $\sigma$-model,
which implies that the couplings $g^i$ and the associated
$\beta$ functions
do not have any explicit scale dependence.
This property of energy conservation can be
extended straightforwardly to many-particle states.
\pr
The same is not true for the angular momentum operator,
in target space dimensions higher than two.
In the context of the target-space effective field theory
this operator is defined as
\be
J^{\alpha \beta } = X^\alpha P_{\beta '}G^{\beta\beta'}
\label{twentyseven}
\ee
where Greek indices denote target spatial components,
and $G_{\alpha\beta}$ is the metric tensor in target space.
The stress tensor is derived from the effective lagrangian
\bea
  T_{\alpha\beta} &=&\frac{\delta {\cal L}^{eff} }{\delta
G_{\alpha\beta}}     \nn \\
T_{0\beta} &=& \frac{\delta {\cal L}^{eff} }{\delta
G_{0\beta}}
\label{twentyeight}
\eea
and the target momenta can be defined from the effective
theory stress tensor by differentiating with respect
to the $G_{0\beta}$ component of the metric (here the ``$0$''
component refers to the Liouville time).
The effective action is identified with the Zamolodchikov
$C$-function. By construction, the latter is
renormalization-group invariant, hence
\be
(\partial _t + \beta^i\partial _i )C=0
\label{twentynine}
\ee
Using the off-shell corollary of the $C$-theorem
$\partial _i C = G_{ij}\beta^j$ \cite{osbornc,mavc}, and
taking into account the fact
that the angular momentum operator
is a functional of $g^i$ only, as seen in equation
(\ref{twentyeight}), we observe that
the $g$-commutator term
in (\ref{twentyfiveb}) vanishes, leaving us with the
following non-trivial result for the temporal
dependence of the angular momentum operator in this framework :
\be
\partial _t  J^{\alpha\beta} = X^{[\alpha }G^{\beta]\gamma}
\frac{\delta [\beta^iG_{ij}\beta^j] }{\delta G_{0\gamma} }
\ne 0
\label{thirty}
\ee
with $[,]$ denoting antisymmetrization. This
expression
is non-zero in general.
\pr
We can evaluate the above expression (\ref{thirty})
in an explicit bosonic string background.
Consider for instance the case of a maximally-symmetric
space-time with a constant dilaton:
\be
R_{MN}=G_{MN} R   \qquad ; \qquad \Phi = const
\label{thirtyb}
\ee
To lowest non-trivial order
in $\alpha '$ the graviton $\beta$-functions are just given by the Ricci
tensor
\be
\beta^G _{MN} = R_{MN} + \dots
\label{thirtyc}
\ee
whilst the quantity
$\beta^i G_{ij} \beta^j$
is given by
\be
 \beta^i G_{ij} \beta^j
  = \frac{3}{16\pi^2} [R + \dots ]
\label{upsilon}
\ee
It should be stressed that in the above formulae
the target manifold includes the Liouville/local renormalization
scale $\phi$ as a time component \cite{emnharc}.
We have ignored for simplicity explicit matter fields,
and concentrated on the gravitational sector. Our formulae
are easily adapted to the more general case with matter deformations.
\pr
It is easy to see that for a maximally-symmetric
non-static universe the result (\ref{thirty})
becomes
\be
   \partial _t J^{\alpha\beta} =\frac{ 3}{8\pi^2}
   X^{[\alpha}\nabla ^{\beta]}
   \partial _t R
\label{thirtyd}
\ee
We can see from (\ref{thirtyd})
that the average $Tr \tilde\rho
\partial _t J^{\alpha\beta} \ne 0$
for time-varying curvatures, e.g., expanding universes \cite{emnharc}
with
$\partial _t R(t)\equiv H(t) R(t)   $, where $H(t)$ is a
Hubble parameter.
For such maximally-symmetric spaces
the following operator relation holds
\be
\partial _t <<J^{\alpha\beta}>>=-\frac{3}{8\pi ^2}
H(t)R <<J^{\alpha\beta}>>
\label{thirtyf}
\ee
showing
a decrease of the average angular momentum
in an expanding universe.
This amounts to a derivation of Mach's principle, analogous
to that
\cite{EO}
in conventional inflationary cosmology.
\pr
We can give a physical interpretation of the above results
by making a direct comparison with the
expanding universe solution in non-critical string theory
of ref. \cite{aben}. It has been shown \cite{polch,mm}
that this model can be directly put in the above framework
of Liouville strings by the identification of the Liouville
field with the target time coordinate $X^0$.
The relevant point here is that in the model of
ref. \cite{aben} in $D >3 $ target-space dimensions
there is a  local antisymmetric
tensor field $B_{MN}=-B_{NM}$,
which is assumed to depend non-trivially on the `cosmic' time
$t\equiv e^{X^0}$, whilst the dilaton is linear in $X^0$.
In string theory  there is an Abelian symmetry
that forces $B_{MN}$ to appear only through its field strength,
$H_{MNP}=\nabla _{[M}B_{NP]}$. For $D=4$ one can define
an axion (pseudoscalar) field $b$ by
\be
    H_{MNP}=e^{\Phi} \epsilon_{MNP\Sigma}\nabla _{\Sigma} b
\label{axion}
\ee
where the dilaton factor is due to scale invariance,
and the time derivative is with respect to $t$.
The axion $b$ may be viewed as
the
Goldstone boson of the target space symmetry $b \rightarrow b + const$
\cite{aben}.
A linear $t  $ dependence in $b$ implies a time-independent
factor in $H_{MNP}$.
This has been interpreted in ref. \cite{aben} as a
signal for spontaneous breaking of Lorentz invariance
(and hence of angular momentum as well).
On the other hand, time-translation invariance is not broken by $b$
and therefore energy is conserved in physical amplitudes derived
from the non-critical string model of ref. \cite{aben}.
In contrast, a constant shift in the dilaton
$\Phi \rightarrow \Phi + const.$  scales the overall
target-space Lagrangian, and hence the Planck constant, and
so the dilaton cannot be viewed as a Goldstone boson of a symmetry
at a quantum level. Time-translation invariance is
formally restored if the correct string vacuum is a
superposition of various ground states corresponding
to different constant values $\Phi _0$ of the dilaton field.
This is the situation indicated by our two-dimensional
toy example considered above. In this
case, since only $s$-wave four-dimensional configurations
are described by this model, the antisymmetric-tensor field
strength
vanishes trivially, and one cannot see explicitly the breaking
of Lorentz invariance. However, even in this case one can
see certain generic features of the above approach.
For instance, as we discusssed in ref. \cite{emnerice},
near the infrared fixed point of the world-sheet
the $\sigma$-model action is described by a `topological
version' of the Wess-Zumino black hole model of
ref. \cite{witt},
which contains a $\theta$ term that is nothing other
than a discrete (topological) antisymmetric tensor background.
This term is essential in yielding instanton deformations
with finite world-sheet action \cite{yung}, and thus capable of breaking
conformal invariance and thereby inducing time flow.
Moreover, the coupling of this $\theta$ term is proportional
to the instanton-renormalized level parameter $k_R (t)$
of the topological Wess-Zumino model \cite{emndollar,emnerice},
related to the integration over topological modes.
The latter contains
logarithmic scale (Liouville time) dependences, which are
assumed
to exponentiate beyond the dilute gas approximation, thereby
leading to an exponential Liouville-dependence
(i.e. linear in the
`cosmic time' $t$)
of the antisymmetric
tensor coupling, exactly as required for a spontaneous breaking
of Lorentz invariance, but not of time-translation invariance.
Moreover, in the two-dimensional black-hole example, a constant shift
in the dilaton field corresponds to a change in the black-hole mass.
Our ground state is assumed to be a foamy superposition of various
microscopic black holes, corresponding to various constant shifts
in the dilaton field. From a world-sheet point
of view, it corresponds to a Kosterlitz-Thouless
plasma (Coulomb gas) of various monopole charges
\cite{emndua}.
When averaging over such states in a quantum theory of gravity,
as we do in (\ref{twentysix}),
time-translation invariance is restored and energy is conserved
on the average,
in agreement with our more general
result stemming from the renormalizability of the underlying
$\sigma$-model.
\pr
It should be stressed that the above physical picture
appears specific to four dimensions.
If it is correct
in a full four-dimensional stringy model of
space-time foam, it
leads to severe restrictions on the Liouville scale-dependence
of the antisymmetric tensor backgrounds.
It would be nice to prove that
this feature is a result of `integrating out' topological (global)
degrees of freedom of the string that cannot be measured
by local scattering experiments, in much the same way as in
the two-dimensional ($s$-wave four-dimensional) example.
At this stage this is only a conjecture, given that a realistic
four-dimensional space-time foamy configuration is not known,
at least to the same level of precision and mathematical
exactness as the
two-dimensional Wess-Zumino black hole case
\cite{witt}.
\pr
Before proceeding, we mention
another important property,
namely the
conservation of total probability:
\be
\partial _t \int dp_idg^i Tr {\tilde \rho } =
\int dg^i dp_i \frac{\partial }{\partial p_i} {\tilde
\rho} G_{ji}\beta^j =0
\label{thirtyone}
\ee
if one assumes that the momentum-boundary terms
vanish. This does not imply that the coupling constant
space has no boundary. On the contrary, it is believed
that the latter is a non-simply connected manifold,
if a topological interpretation in terms of Morse theory
is to be given \cite{vafa}.
\pr
\section{Application to the Neutral Kaon System}
\pr
We are now well equipped to discuss the phenomenology
of violations of quantum mechanics in the above
framework. The formalism of \cite{ehns} will be adopted.
Below, we describe briefly the formalism \cite{ehns,emncpt}
for a discussion of the
possible modification of quantum mechanics and violation of
$CPT$ in the neutral kaon system, which is among the most
sensitive microscopic laboratories for studying these
possibilities. In the normal quantum-mechanical
formalism, the time-evolution of a neutral kaon
density matrix is given by
\be
\partial _t \rho = -i (H\rho - \rho H^{\dagger})
\label{cpteight}
\ee
\nk where the Hamiltonian takes the following form
in the ${(K^0 , {\overline K}^0 )}$ basis:
\be
  H = \left( \begin{array}{c}
 (M + \frac{1}{2}\Delta M) - \frac{1}{2}i(\Gamma + \frac{1}{2}
 \Delta \Gamma)
   \qquad  \qquad
   M_{12}^{*} - \frac{1}{2}i\Gamma _{12}^{*} \\
           M_{12}  - \frac{1}{2}i\Gamma _{12}
    \qquad  \qquad
    (M - \frac{1}{2}
    \Delta M)-\frac{1}{2}i(\Gamma
    - \frac{1}{2}
    \Delta \Gamma ) \end{array}\right)
\label{cptnine}
\ee
The non-hermiticity of $H$ reflects the process of
$K$ decay: an initially-pure state evolving
according to (\ref{cpteight}) and (\ref{cptnine})
remains pure.
\pr
In order to discuss the possible modification of this
normal quantum-mechanical evolution, and allow for the
possibility of $CPT$ violation, it is convenient to
rewrite \cite{emncpt}
(\ref{cpteight}) and (\ref{cptnine}) in a
Pauli $\sigma$-matrix basis \cite{ehns}, introducing
components ${\rho}_{\alpha}$ of the density matrix:
\be
\rho = 1/2 \rho_{\alpha} \sigma_{\alpha}
\label{cptten}
\ee
which evolves according to
\be
 \partial _t \rho_\alpha  =
h_{{\alpha}{\beta}}{\rho_{\beta}}
\label{cpteleven}
\ee
with
\be
  h_{\alpha\beta} \equiv \left( \begin{array}{c}
 Imh_0 \qquad Imh_1 \qquad Imh_2 \qquad Imh_3 \\
 Imh_1 \qquad Imh_0 \qquad -Reh_3 \qquad Reh_2 \\
 Imh_2 \qquad Reh_3 \qquad Imh_0 \qquad -Reh_1 \\
 Imh_3 \qquad -Reh_2 \qquad Reh_1 \qquad Imh_0 \end{array}\right)
\label{cptelevenb}
\ee
It is easy to check that at large times $\rho$ takes the form
\be
  \rho \simeq e^{-\Gamma _Lt}
 \left( \begin{array}{c}
 1   \qquad  \epsilon^*            \\
 \epsilon          \qquad |\epsilon |^2 \end{array}\right)
\label{cpttwelve}
\ee
where $\epsilon$ is given by
\be
  \epsilon =\frac{\frac{1}{2}i Im \Gamma _{12} - Im M_{12}}
{\frac{1}{2} \Delta \Gamma - i\Delta M }
\label{cptthirteen}
\ee
in the usual way.
\pr
A modification of quantum mechanics of the form
discussed in section 3 can be introduced by modifying
equation (\ref{cpteleven}) to become
\be
\partial _t \rho_\alpha  = h_{\alpha\beta} \rho_\beta +
\nd{h}_{\alpha\beta}\rho_\beta
\label{cptfourteen}
\ee
The form of ${\nd h}_{\alpha \beta}$ is determined if we
assume probability and energy conservation, as proved
in the string context in section 3, and that the
leading modification conserves strangeness:
\be
  {\nd h}_{\alpha\beta} =\left( \begin{array}{c}
 0  \qquad  0 \qquad 0 \qquad 0 \\
 0  \qquad  0 \qquad 0 \qquad 0 \\
 0  \qquad  0 \qquad -2\alpha  \qquad -2\beta \\
 0  \qquad  0 \qquad -2\beta \qquad -2\gamma \end{array}\right)
\label{cptfifteen}
\ee
It is easy to solve the $4 \times 4$ linear matrix equation
(\ref{cptfourteen}) in the limits of large time:
\be
\rho _L
\propto \left( \begin{array}{c} 1 \qquad  \qquad
\frac{-\frac{1}{2}i  (Im \Gamma _{12} + 2\beta )- Im M_{12} }
{\frac{1}{2} \Delta \Gamma + i \Delta M } \\
\frac{\frac{1}{2}i (Im \Gamma _{12} + 2\beta )- Im M_{12} }
{\frac{1}{2}\Delta \Gamma  - i \Delta M} \qquad \qquad
|\epsilon |^2 + \frac{\gamma}{\Delta \Gamma } -
\frac{4\beta Im M_{12} (\Delta M / \Delta \Gamma ) + \beta ^2 }
{\frac{1}{4} \Delta \Gamma ^2 + \Delta M^2 } \end{array} \right)
\label{cptsixteen}
\ee
and of short time:
\be
 \rho _S
 \propto \left( \begin{array}{c} |\epsilon |^2 +
\frac{\gamma }{|\Delta \Gamma |} -
\frac{-4\beta Im M_{12} (\Delta M/\Delta \Gamma )+ \beta ^2 }
{ \frac{1}{4} \Delta \Gamma ^2 + \Delta M^2 } \qquad
\epsilon - \frac{i\beta}{\frac{\Delta \Gamma} {2} - i \Delta M} \\
\epsilon ^* + \frac{i\beta} {\frac{\Delta \Gamma}{2} +
i \Delta M} \qquad 1 \end{array} \right)
\label{cptseventeen}
\ee
We note that the density matrix (\ref{cptsixteen}) for
$K_L$ is mixed to the extent that the parameters $\beta$
and $\gamma$ are non-zero. It is also easy to check \cite{emncpt}
that
the parameters $\alpha$, $\beta$ and $\gamma$ all violate
$CPT$, in accord with the general argument of \cite{wald},
and consistent with the string analysis mentioned earlier
in this section.
\pr
Experimental observables $O$ can be introduced \cite{ehns,emncpt}
into this
framework as matrices, with their measured values being given by
\be
<O> = Tr(O \rho)
\label{cpteighteen}
\ee
Examples are the $K$ to $2 \pi$ and $3 \pi$ decay
observables
\be
 O_{2\pi} =\left( \begin{array}{c} 0 \qquad 0 \\
0 \qquad 1 \end{array} \right)
\qquad ; \qquad
 O_{3\pi}
 =(0.22)
 \left( \begin{array}{c} 1 \qquad 0 \\
0 \qquad 0 \end{array} \right)
\label{cptnineteen}
\ee
and the semileptonic decay observables
\bea
  O_{\pi ^-l^+ \nu} = \left(\begin{array}{c} 1 \qquad 1 \\
1\qquad 1 \end{array}\right)  \nn \\
  O_{\pi ^+l^-{\overline \nu}}  =\left( \begin{array}{c}
1 \qquad -1 \\
 -1\qquad 1 \end{array} \right)
\label{cpttwenty}
\eea
A quantity of interest is the difference between
the $K_L$ to $2 \pi$ and $K_S$ to $3 \pi$
decay rates \cite{emncpt}:
\be
\delta R \equiv R_{2\pi}-R_{3\pi}= \frac{8\beta}{|\Delta \Gamma |}
 |\epsilon| sin\phi _{\epsilon}
\label{cpttwentyhalf}
\ee
where
$R_{2\pi}^L\equiv Tr(O_{2\pi} \rho_L)$,
and $R_{3\pi}^S \equiv Tr(O_{3\pi}\rho_S)/0.22 $, and the prefactors
are determined by the measured \cite{pdg}
branching ratio for
$K_L \rightarrow 3{\pi}^0$. (Strictly speaking, there should be a
corresponding prefactor of $0.998$ in the formula
(\ref{cptnineteen}) for
the $O_{2{\pi}}$ observable.)
\pr
Using (\ref{cpttwenty}), one can calculate the semileptonic
decay asymmetry \cite{emncpt}
\be
\delta \equiv \frac{\Gamma (\pi^-l^+\nu) - \Gamma (\pi^+ l^-
{\overline \nu }) }{\Gamma (\pi ^- l^+ \nu ) +
\Gamma (\pi ^+ l^- {\overline \nu})}
\label{cpttwentyone}
\ee
in the long- and short-lifetime limits:
\bea
\delta_L  & = &   2Re [\epsilon (1-\frac{i\beta}{Im M_{12}})]   \nn \\
       \delta_S   & = &
2Re [ \epsilon (1 + \frac{i\beta}{Im M_{12}})]
\label{cpttwentytwo}
\eea
The difference between these two values
\be
          \delta\delta \equiv
\delta _L  - \delta _S  = -\frac{8\beta} {|\Delta \Gamma |}
\frac{sin\phi _{\epsilon}}{\sqrt{1 + \tan^2 \phi _{\epsilon}}}
= -\frac{8\beta}
{|\Delta \Gamma |} sin\phi _{\epsilon}
cos\phi _{\epsilon}
\label{cpttwentythree}
\ee
with $tan\phi _\epsilon =(2\Delta M)/\Delta \Gamma $,
is a signature
of $CPT$ violation that can be explored at the
CPLEAR and DA$\phi$NE facilities \cite{cplear,dafne}.
\pr
We have used \cite{emncpt} the latest experimental values of
$R_{2 \pi}$ and $R_{3 \pi}$ to bound $\delta R$,
and the latest experimental values of $\delta_{L,S}$
to bound $\delta \delta$, expressing the results as
contours in the $(\beta , \gamma )$ plane \cite{emncpt}.
In our formalism, the usual
$CP$-violating parameter $\epsilon$ is given
by \cite{emncpt}
\be
|\epsilon| = -\frac{2\beta}{|\Delta \Gamma |}sin\phi _{\epsilon}
+ \sqrt{\frac{4\beta ^2}{|\Delta \Gamma |^2 } - \frac{\gamma}
{|\Delta \Gamma |} + R_{2\pi}^L }
\label{cpttwentyfour}
\ee
On the basis of this preliminary analysis, it is safe
to conclude that
\be
|\frac{\beta}{\Delta \Gamma}| \lsim ~10^{-4}~to~10^{-3}
\qquad ; \qquad
|\frac{\gamma}{\Delta \Gamma}| \lsim ~10^{-6}~to~10^{-5}
\label{cpttwentyfive}
\ee
In addition to more precise experimental data,
what is also needed is a more complete global fit
to all the available experimental data, including
those at intermediate times, which are essential for
bounding $\alpha$, and may improve our bounds
(\ref{cpttwentyfive}) on $\beta$ and $\gamma$
\cite{emncpt,elmn,huet}.
We now give a brief account of the
intermediate-time formalism.
As a first step,
we consider a perturbative ansatz for
the density matrix elements $\rho _{ij}$, $i,j = 1,2$ that
appear in the system of equations (\ref{cptfourteen}) after
a (convenient)
change of basis to $K_{1,2} \equiv \sqrt{1}{\sqrt{2}}(K^0 \mp
{\overline K}^0 )$ \cite{ehns,emncpt}. We write
\cite{elmn}
\be
\rho_{ij} (t) = \rho _{ij}^{(0)} + \rho _{ij} ^{(1)} + \dots
\label{rhopert}
\ee
where $\rho_{ij}^{(k)}$ are polynomials  in $\alpha$, $\beta$,
$\gamma$ and $|\epsilon|$ of degree $k$ :
\be
\rho _{ij}^{(k)} \equiv \alpha ^{P_\alpha} \beta ^{P_\beta}
\gamma ^{P_\gamma} |\epsilon |^{P_\epsilon}
\qquad ; \qquad P_\alpha + P_\beta + P_\gamma + P_\epsilon = k
\label{polyn}
\ee
with the initial condition of having a pure $K^0$ state, i.e.,
$\rho _{ij}^{0} (0)=\frac{1}{2}$,
$\rho ^{k}_{ij} (0) = 0$, $k \ge 1 $. The ansatz
(\ref{rhopert}) leads to the following iterative system
of differential equations, describing the time evolution
of the density matrix of the neutral-kaon system at arbitrary time
intervals \cite{elmn} :
\be
\frac{d}{d t} [e^{A t} \rho _{ij} ^{(k)} (t) ] = e^{A t}
\sum _{kl \ne ij} \rho ^{(k-1)}_{kl} (t)
\label{pertsyst}
\ee
where $A$ is a generic factor that
can be expressed in terms of known data of the
neutral-kaon system \cite{elmn}.
In the long and short time limits one recovers the bounds
(\ref{cpttwentyfive})
of $\beta $ and $\gamma$ . On the other hand,
a fit to presently available intermediate
time data
from two-pion decays \cite{gibbons}
can place more stringent bounds  on these quantities \cite{elmn},
confirming that
the
standard $CP$-violation
(\ref{cpttwentyfour}),
observed so far, is mainly quantum mechanical in origin.
Moreover, an upper bound
on the quantity
$\alpha$ can also be placed by such fits,
\be
 |\frac{\alpha}{\Delta \Gamma}| \lappeq 2 \times 10^{-3}
\label{alphabound}
\ee
although more stringent bounds can be placed by a study of
$\phi$-decays at a $\phi$-factory \cite{dafne,huetalk,huet}.
\pr
A concrete phenomenological
consequence of the $CPT$-violation will be a shift $\delta \phi $
in the minimum
of the time-dependent semileptonic decay asymmetry $\delta (t)$
(\ref{cpttwentyone}) as a function of time $t$. A preliminary
estimate  of this shift, using the bound (\ref{alphabound})
yields \cite{elmn}
\be
\delta \phi \lappeq 6 \times 10^{-3}
\label{shift}
\ee
and we expect this range to be probed in the foreseeable future.
\pr
We cannot resist pointing out that the bounds
(\ref{cpttwentyfive}) are quite close to
\be
O(\Lambda_{QCD} / M_P ) m_K \simeq 10^{-19} GeV
\label{cpttwentysix}
\ee
which is perhaps the largest magnitude that any
such $CPT$- and quantum-mechanics-violating
parameters could conceivably have. Since any
such effects are associated with topological
string states that have masses of order $M_P$,
we expect them to be suppressed by some power of
$1/M_P$. This expectation is supported by the
analogy with the Feynman-Vernon model of
quantum friction \cite{vernon}, in which coherence is
suppressed by some power of the unobserved
oscillator mass or frequency. If the $CPT$- and
quantum-mechanics-violating parameters discussed in
this section are suppressed by just one power of
$M_P$, they may be accessible to the next round of
experiments with CPLEAR and/or DA$\phi$NE \cite{cplear,dafne}.
\pr
\pr
\nk {\Large{\bf  Acknowledgements}}
\pr
The work of N.E.M. is supported by a EC Research Fellowship,
Proposal Nr. ERB4001GT922259 .
That of D.V.N. is partially supported by DOE grant
DE-FG05-91-GR-40633.
\vfil\eject

\pr
\newpage
{\Large {\bf Figure Captions}}
\pr
\nk {\bf Figure 1 } - Contour
of integration in the analytically-continued
(regularized) version of $\Gamma (-s)$ for $ s \in Z^+$.
This is known in the literature as the Saalschutz contour,
and has been used in
conventional quantum field theory to relate dimensional
regularization to the Bogoliubov-Parasiuk-Hepp-Zimmermann
renormalization method.
\pr
\nk {\bf Figure 2 } - Schematic repesentation
of the evolution of the world-sheet area as the renormalization
group scale moves along the contour of fig. 1.
\pr


\begin{thebibliography}{99}
\bibitem{mandelstam} S. Mandelstam, Phys. Lett. B277 (1992), 82.
\bibitem{bek} S. Hawking, Comm. Math. Phys. 43 (1975), 199 ;
\par J. Bekenstein, Phys. Rev. D12 (1975), 3077.
\bibitem{hawk} S. Hawking, Comm. Math. Phys. 87 (1982), 395.
\bibitem{ehns} J. Ellis, J.S. Hagelin, D.V. Nanopoulos and
M. Srednicki, Nucl. Phys. B241 (1984), 381.
\bibitem{emohn} J. Ellis, S. Mohanty and D.V. Nanopoulos,
Phys. Lett. B221 (1989), 113.
\bibitem{emnqm} J. Ellis, N.E. Mavromatos
and D.V. Nanopoulos, Phys. Lett. B293 (1992), 37.
\bibitem{emndollar} J. Ellis, N.E. Mavromatos and
D.V. Nanopoulos, CERN, ENS-LAPP and Texas A \& M Univ. preprint,
CERN-TH.6896/93, ENS-LAPP-A426-93, CTP-TAMU-29/93; ACT-09/93 (1993);
hep-th/9305116 ;
\par J. Ellis, N.E. Mavromatos and
D.V. Nanopoulos, CERN, ENS-LAPP and Texas A \& M Univ. preprint,
CERN-TH.6897/93, ENS-LAPP-A427-93, CTP-TAMU-30/93; ACT-10/93 (1993);
hep-th/9305117.
\bibitem{aben} I. Antoniadis, C. Bachas, J. Ellis
and D.V. Nanopoulos, Phys. Lett. B211 (1988), 393;
Nucl. Phys. B328 (1989), 117; Phys. Lett. B257 (1991), 278.
\bibitem{polch} J. Polchinski, Nucl. Phys. B324 (1989), 123;
\par D.V. Nanopoulos, in {\it Proc. Int. School
of Astroparticle Physics}, HARC, Houston (World Scientific, Singapore,
1991),
p. 183.
\bibitem{emnerice} J. Ellis, N.E. Mavromatos
and D.V. Nanopoulos, preprint CERN-TH.7195/94, ENS-LAPP-A-463/94,
ACT-5/94, CTP-TAMU-13/94, {\it lectures presented at the
Erice Summer School, 31st Course: From Supersymmetry to the
Origin of Space-Time},
Ettore Majorana Centre, Erice, July 4-12
1993.
\bibitem{emnharc} J. Ellis, N.E. Mavromatos
and D.V. Nanopoulos,
{\it Proc. HARC workshop on ``Recent Advances in the
Superworld", The Woodlands, Texas (USA), April 14-16 1993},
eds. J. Lopez and D.V. Nanopoulos, (World Sci., Singapore,
1994),p. 3;
preprint CERN-TH.7000/93, CTP-TAMU 66/93,
ENSLAPP-A-445/93, OUTP-93-26P, hep-th/9311148,
\bibitem{zam} A.B. Zamolodchikov, JETP Lett. 43 (1986), 730;
Sov. J. Nucl. Phys. 46 (1987), 1090.
\bibitem{emn1} J. Ellis, N.E.  Mavromatos and
D.V. Nanopoulos,
Phys. Lett. B267 (1991), 465; {\it ibid}
B272 (1991), 261.
\bibitem{pen} R. Penrose, {\it The Emperor's New Mind}
(Oxford Univ. Press, 1989).
\bibitem{emncpt} J. Ellis, N.E. Mavromatos and
D.V. Nanopoulos, Phys. Lett. B293 (1992), 142;
CERN and Texas A \& M Univ. preprint
CERN-TH. 6755/92; ACT-24/92;CTP-TAMU-83/92; hep-th/9212057 ;
\bibitem{wald} R. Wald, Phys. Rev. D21 (1980), 2742.
\bibitem{elmn} J. Ellis, J. Lopez, N.E. Mavromatos
and D.V. Nanopoulos, in preparation ;
\par J. Lopez, preprint
CPT-TAMU 38/93 (1993), Proc.
at {\it HARC workshop on ``Recent Advances in the
Superworld", The Woodlands, Texas (USA), April 14-16 1993},
eds. J. Lopez and D.V. Nanopoulos, (World Sci., Singapore, 1994),
p.~272.
\bibitem{huetalk} P. Huet, these proceedings.
\bibitem{huet} P. Huet and M. Peskin, SLAC (Stanford) preprint
SLAC-PUB-6454 (1994), hep-ph/9403257.
\bibitem{mp} B. Misra, I. Prigogine and M. Courbage,
{\it Physica} A98 (1979), 1;
\par I. Prigogine, {\it Entropy, Time, and
Kinetic Description}, in {\it Order and Fluctuations
in Equilibrium and Non-Equilibrium Statistical Mechanics},
ed G. Nicolis et al. (Wiley, New York, 1981);
\par B. Misra and I. Prigogine, {\it Time, Probability and Dynamics},
in {\it Long-Time Prediction in Dynamics}, ed G. W. Horton,
L. E. Reichl and A.G. Szebehely (Wiley, New York, 1983) ;
\par B. Misra, {\it Proc. Nat. Acad. Sci. U.S.A.}
75 (1978), 1627.
\bibitem{ktorides} J.P. Constantopoulos and
C.N. Ktorides, J. Phys. A17 (1984), L29.
\bibitem{fronteau} J. Fronteau, A. Tellez-Arenas
and R.M. Santilli, Hadronic J. 3 (1979), 130;
\par J. Fronteau, Hadronic J. 4 (1981), 742.
\bibitem{wittop} D. Gross, Phys. Rev. Lett. 60 (1988), 1229;
\par E. Witten, Trans. Roy. Soc. A329 (1989), 345.
\bibitem{emntop} J. Ellis, N.E. Mavromatos and
D.V. Nanopoulos, Phys. Lett. B288 (1992), 23.
\bibitem{shore} G. Shore, Nucl. Phys. B286 (1987), 349.
\bibitem{osborn} H. Osborn, Nucl. Phys.
B294 (1987), 595; {\it ibid}
B308 (1988), 629; Phys. Lett. B222 (1989), 97.
\bibitem{witt} E. Witten, Phys. Rev. D44 (1991), 314.
\bibitem{polybook} A.M. Polyakov, {\it Gauge
Fields and Strings} (Harwood, New York, 1987).
\bibitem{stringbook} M.B. Green, J.H. Schwarz and E. Witten,
{\it String Theory}, Vol. I and II (Cambridge Univ. Press, 1986).
\bibitem{fischler} W. Fischler and L. Susskind, Phys.
Lett. B171 (1986), 262; {\it ibid} B171 (1986), 383.
\bibitem{pol} A.M. Polyakov, Mod. Phys. Lett. A6 (1991), 635.
\bibitem{DDK}F. David, Mod. Phys. Lett. A3 (1988), 1651;
\par J. Distler and H. Kawai, Nucl. Phys. B321 (1989), 509.
\bibitem{mm} N.E. Mavromatos and J.L. Miramontes,
Mod. Phys. Lett. A4 (1989), 1847.
\bibitem{frolov} V. Frolov, {\it Physics of 2-D Black Holes},
Proc. {\it Workshop on Strings and Quantum Gravity},
Ettore Majorana Centre, Erice,
June 1992, Eds. N. Sanchez and A. Zichichi, (World Sci., Singapore,
1993)
\bibitem{chlyk} S. Chaudhuri and J. Lykken, Nucl. Phys B396 (1993),
270.
\bibitem{continu} J. Distler and P. Nelson, Nucl. Phys. B374 (1992),
123.
\bibitem{emnhair} J. Ellis, N.E. Mavromatos and
D.V. Nanopoulos, Phys. Lett. B284 (1992), 43.
\bibitem{emn4d} J. Ellis, N. Mavromatos and D.V. Nanopoulos,
Phys. Lett. B278 (1992), 246.
\bibitem{venez} G. Veneziano, Phys. Lett. B167 (1985), 388;
\par J. Maharana and G. Veneziano, Nucl. Phys. B283 (1987), 126;
\par T. Kubota, Osaka Univ. preprint OU-HET-146 (1990),
Proc.
{\it Symposium on
Quarks, Symmetries and Strings, in honour of the
60th birthday
of B. Sakita, (New York, October 1-2, 1990)}.
\bibitem{evans} M. Evans and B. Ovrut, Phys. Rev. D39 (1989), 3016.
\bibitem{ovrut} M. Evans and B. Ovrut, Phys. Rev. D41 (1990), 3149.
\bibitem{bakir} I. Bakas and E. Kiritsis, Phys. Lett. B301 (1993),
49.
\bibitem{evovr} M. Evans and B. Ovrut, Phys. Lett. B231 (1989), 80.
\bibitem{coleman} S. Coleman, Comm. Math. Phys. 31 (1973), 253.
\bibitem{witt2} E. Witten, Nucl. Phys. B373 (1992), 187.
\bibitem{curci} G. Curci and G. Paffuti, Nucl. Phys. B312 (1989), 227.
\bibitem{mavc} N.E. Mavromatos and J.L. Miramontes,
Phys. Lett. B212 (1988), 33;
\par N.E. Mavromatos, Phys. Rev. D39 (1989), 1659.
\bibitem{espriu} D. Espriu and N.E. Mavromatos, Phys. Lett. B237
(1990), 370.
\bibitem{matrix} E. Br\'ezin and V.A. Kazakov,
Phys. Lett. B236 (1990), 144;
\par M. Douglas and A. Shenker, Nucl. Phys.
B335 (1990), 635;
\par D. Gross and A.A. Migdal, Phys. Rev. Lett.
64 (1990), 127;
\par For a recent
review see, e.g., I. Klebanov,
in {\it String Theory and Quantum Gravity}, Proc. Trieste
Spring School 1991, ed. by J. Harvey et al.
(World Scientific, Singapore, 1991), and references therein.
\bibitem{bhmatrix} A. Jevicki and T. Yoneya, Nucl. Phys.
B411 (1994), 64.
\bibitem{eguchi} T. Eguchi, Mod. Phys. Lett. A7 (1992), 85.
\bibitem{mukhi} S. Mukhi and C. Vafa, Harvard Univ. and
Tata Inst. preprint HUTP-93/A002; TIFR/TH/93-01 (1993).
\bibitem{kutasov} D. Kutasov, Mod. Phys. Lett.
A7 (1992), 2943.
\bibitem{kogan} I. Kogan, Phys. Lett. B265 (1991), 269.
\bibitem{Li} M. Goulian and M. Li, Phys. Rev. Lett.
66 (1991), 2051.
\bibitem{bershadsky} M. Bershadsky and D. Kutasov, Phys. Lett.
B266 (1991), 345.
\bibitem{partovi} H. Partovi, Phys. Rev. Lett. 50 (1983), 1885;
\par R. Blankenbecler and H. Partovi, Phys. Rev. Lett.
57 (1986), 2887.
\bibitem{zurek} W. Zurek, S. Habib and J.P. Paz, Phys.
Rev. Lett. 70 (1992), 1187.
\bibitem{albrecht} A. Albrecht, Phys. Rev. D46 (1992), 5504.
\bibitem{barton} G. Barton, Ann. Phys. 166 (1986), 322.
\bibitem{cald} A.O. Caldeira and A.J. Leggett, Ann. Phys.
149 (1983), 374.
\bibitem{hu} B.L. Hu and Y.H. Zhang, Univ. of Maryland prperint,
UMD-PP-93-161 (1993)/gr-qc bulletin 930811,
and references therein.
\bibitem{thomas} B.A. Ovrut and S. Thomas, Phys. Rev. D43 (1991),
1314.
\bibitem{ciaf} D. Amati, M. Ciafaloni and G. Veneziano,
Phys. Lett. B197 (1987), 81; {\it ibid} B216 (1989), 41.
\bibitem{elze} T Elze, preprint CERN-TH.7131/93 (1994);
hep-ph/9404215.
\bibitem{emndua} J. Ellis, N.E. Mavromatos and D.V. Nanopoulos,
Phys. Lett. B289 (1992), 25; {\it ibid} B296 (1992), 40.
\bibitem{pruisk} A. Pruisken, Nucl. Phys. B290[FS20] (1987),
61.
\bibitem{cplear} R. Adler {\it et al.}, CPLEAR-Collaboration,
{\it A study of T violation via the semileptonic
Decays of Neutral Kaons in CPLEAR}, paper submitted
to {\it Second Biennial Conference on Low-Energy Antiproton
Physics (LEAP), Courmayeur 1992};
\bibitem{dafne} {\it Da$\phi$ne Physics Handbook}, edited by
L. Maiani, L. Pancheri and N. Paver (INFN, Frascati, 1992).
\bibitem{osbornc} H. Osborn, Phys. Lett. B222 (1989), 97.
\bibitem{EO} J. Ellis and K.A. Olive, {\it Nature} {\bf 303}
(1983) 679.
\bibitem{yung} A.V. Yung,
Int. J. Mod. Phys. A9 (1994), 591 ;
\par A.V. Yung, Swansea preprint SWAT 94/22
(1994).
\bibitem{vafa} C. Vafa, Phys. Lett. B212 (1988), 28.
\bibitem{pdg} {\it Review of Particle Properties},
Particle Data Group, Phys. Rev. D45 (No 11, part II) (1992), 1.
\bibitem{gibbons} L.K. Gibbons, {\it et al.}, E731-Collaboration,
Phys. Rev. Lett. 70 (1993), 1199 and references therein.
\bibitem{vernon} R.P. Feynman and F.L. Vernon Jr., Ann. Phys.
(NY) 94 (1963), 118.
\end{thebibliography}
\end{document}